\documentclass[11pt]{article}

\usepackage{axodraw}
\usepackage{epsf}
\usepackage[usenames]{color}
\usepackage{epsfig}
\usepackage{graphicx}
\usepackage{bbm}
\usepackage{amsmath}
\usepackage{amssymb}
\usepackage{amsfonts}
\usepackage{mathrsfs}
\usepackage{cite}
\newcommand{\bc}{\begin{center}}
\newcommand{\ec}{\end{center}}
\newcommand{\bt}{\begin{tabular}}
\newcommand{\et}{\end{tabular}}
\newcommand{\be}{\begin{equation}}
\newcommand{\ee}{\end{equation}}
\newcommand{\bea}{\begin{eqnarray}}
\newcommand{\eea}{\end{eqnarray}}
\newcommand{\bfig}{\begin{figure}}
\newcommand{\efig}{\end{figure}}
\def\gsim{ \lower .75ex \hbox{$\sim$} \llap{\raise .27ex \hbox{$>$}} }
\def\lsim{ \lower .75ex \hbox{$\sim$} \llap{\raise .27ex \hbox{$<$}} }

\begin{document}
\setlength{\unitlength}{1mm}
{\vspace*{-2cm}
\flushright{ULB-TH/09-03}\\
\vspace*{-0.5cm}
\flushright{FTUAM-09-04}
\vspace{-4mm}
\vskip 1.5cm}
\bc
{\Large \bf Scalar Multiplet Dark Matter}
\ec
\vskip 0.5cm
\bc
{\large T.~Hambye$^{\hbox{a}}$, F.-S. Ling$^{\hbox{a}}$, L.~Lopez Honorez$^{\hbox{a},\hbox{b}}$ and J.~Rocher$^{\hbox{a}}$
\footnote{thambye@ulb.ac.be; fling@ulb.ac.be; llopezho@ulb.ac.be; jrocher@ulb.ac.be}}
\\
\vskip .5cm
$^{\hbox{a}}$Service de Physique Th\'eorique, Universit\'e Libre de Bruxelles, 1050 Brussels, Belgium\\
\vspace{0.7mm}
$^{\hbox{b}}$Depto de Fisica Teorica, Universidad Autonoma de Madrid, Cantoblanco, Madrid, Spain\\
\ec
\vskip 0.5cm
\begin{abstract}
  We perform a systematic study of the phenomenology associated
 to models where the dark matter consists in the neutral component of a scalar
 $SU(2)_L$ $n$-uplet, up to $n=7$. If one includes only the pure gauge induced
 annihilation cross-sections it is known that such particles provide good dark
 matter candidates, leading to the observed dark matter relic 
abundance for a particular value of their mass around the TeV scale. 
We show that these values actually become ranges of values - which we
determine - if one takes into account the annihilations induced by the various
 scalar couplings appearing in these models. This leads to predictions for
 both direct and indirect detection signatures as a function of the dark matter
 mass within these ranges. Both can be largely enhanced by the quartic
 coupling contributions.
We also explain how, if one adds right-handed neutrinos to the scalar doublet
case, the results of this analysis allow to have altogether a viable
 dark matter candidate, successful generation of neutrino masses, 
and leptogenesis in a particularly minimal way with all new physics at the TeV scale.

\end{abstract}

\setcounter{footnote}{0}

\section{Introduction}

There are many possible dark matter (DM) candidates and a systematic study of all 
possibilities, and associated phenomenology, is not conceivable. However if one takes as 
criteria the minimality of the model, in terms of the number of new fields and parameters, such a systematic 
study becomes feasible. 
Such approach is different and complementary  to the ones that led to theories with {\it e.g.}
Supersymmetry~\cite{Goldberg:1983nd,Ellis:1983ew,Jungman:1995df} or 
Universal Extra Dimensions~\cite{Servant:2002aq,Cheng:2002ej,Hooper:2007qk}, which were invented as an attempt to address and solve other fundamental questions 
such as the Hierarchy problem,
and where the number of parameters and possibilities can be huge. 
Another criterion of selection one can consider is the predictivity and the testability of the model in current and future
accelerators, and direct or indirect DM detection experiments.
Particularly simple possibilities along these lines of thought arise if one adds to the Standard Model 
only one extra $SU(2)_L$ singlet or multiplet, 
scalar or fermion, containing a neutral DM candidate field. 
The stability of the DM is usually achieved
in this case by introducing a $Z_2$ parity symmetry, under which the extra multiplet is odd and all the SM particles 
are even. 

Several possibilities of this kind, such as 
the scalar singlet \cite{McDonald:2001vt,Burgess:2000yq,Boehm:2003hm,Boehm:2008zz,Espinosa:2008kw,Barger:2007im,Andreas:2008xy,Yaguna:2008hd}, 
the fermion singlet \cite{Pospelov:2007mp,Belanger:2007dx}, 
the scalar doublet (in the "Inert Doublet Model"~\cite{Deshpande:1977rw,Ma:2006km,Barbieri:2006dq,Majumdar:2006nt,LopezHonorez:2006gr,Gustafsson:2007pc,Hambye:2007vf,Andreas:2008xy,Kadastik:2009dj}), 
the fermion doublet candidate \cite{Cirelli:2005uq,Essig:2007az}, etc, 
 have already been explored and they offer a rich phenomenology.
A systematic study has been performed in Ref.~\cite{Cirelli:2005uq} for any multiplet from the doublet up to the 
7-plet. Multiplets offer the advantage that they could be potentially produced at colliders through gauge interactions. 
In this analysis the relic density of such DM candidate has been calculated 
considering all annihilation processes induced by the known $SU(2)_L\times U(1)_Y$ gauge 
interactions. This framework is particularly predictive, the only free
parameter is the DM mass, $m_{DM}$, and the observed relic density can be
obtained  for only one value of this mass.

Considering only the gauge induced processes in such a way is fully justified for an 
extra fermion multiplet because no other renormalizable interaction with the SM particles can be written. 
However, for a scalar multiplet this assumption is not at all automatic as quartic scalar interactions 
involving both the scalar multiplet and the Brout-Englert-Higgs doublet are perfectly allowed.
Therefore, the analysis of Ref.~\cite{Cirelli:2005uq} for scalar multiplets does not hold if these scalar couplings are not suppressed. 

In this paper, we study in a systematic way the rich phenomenology which arise if one
includes the effects of the quartic couplings for all scalar multiplets up to the 7-plet. This we do
in the high mass regime, that is to say for $m_{DM} > m_W$, where the observed relic density is obtained 
for annihilation cross-sections $\propto 1/m_{DM}^2$ (which is typical of the large DM mass asymptotic regime).
We show in particular that, due to a large \emph{enhancement} of 
the (co)annihilation of DM into gauge bosons driven by the 
scalar couplings, the latter cannot be ignored unless they are much smaller than the gauge couplings. 
Moreover, due to these quartic couplings, and without fine-tuning, a large
range of values of $m_{DM}$ 
is compatible with the observed DM relic abundance.
These contributions also enhances the predicted fluxes for direct and indirect detection searches. 

The case where the multiplet is a doublet, known as the Inert Doublet Model (IDM)
has already been extensively studied in the literature.
A detailed analysis of the high mass regime was however missing and we provide it here. 
The phenomenology of this model is particularly rich because it depends on the interplay of three different
scalar quartic couplings.
The phenomenology of the higher multiplet case on the other hand {\it in fine} depends 
on only one quartic coupling,  $\lambda_3$, which renders these cases particularly constrained and predictive. 
In this work, only higher multiplet models allowed by current direct detection 
constraints will be considered. This limits us to odd dimension $n$-uplets with zero hypercharge
and $n=3,5$~and~$7$. For these models the high mass regime, $m_{DM} > m_W$, which we study is the only possible one.

We also present in this paper an intriguing possible consequence of our results for the doublet model: in agreement with the DM constraints, if,
in order to explain the neutrino masses,  one adds to this model right-handed neutrinos, 
it is possible to induce in a particularly simple way baryogenesis through leptogenesis with all new physics around the TeV scale.

The paper is organized as follows. We first present the inert doublet and the higher multiplet models
in Section~\ref{sec:models}. Aside from fixing notations and definitions, 
a discussion on the number of relevant quartic couplings for the higher multiplets is made.
Predictions for the relic density are made in Section~\ref{sec:relic}, both 
numerically and (in the instantaneous freeze-out approximation) analytically. The latter method allows to
show the enhancement of the scalar coupling contribution in the various cross-sections, in particular the important coannihilation ones.
For the doublet case, maximal mass splittings between the DM doublet components compatible
with the WMAP constraint are given as a function of the DM mass.
For higher multiplets, this constraint fixes the value of $\lambda_3$ as a function of the DM mass.
A discussion is also made about the consequences of having, for very heavy DM candidates, freeze-out before the electroweak phase transition.
In Section~\ref{sec:DD}, predictions on the DM-nucleon elastic scattering cross-section relevant for direct
detection searches are made, and compared to current experimental limits and projected reaches
of future experiments. 
In Section~\ref{sec:ID}, predictions for various indirect detection signals are discussed.
The possibility of resonances~\cite{Hisano:2006nn,Cirelli:2007xd} is reexamined in light of the enlarged mass range
of the DM candidate. 
Photon and neutrino fluxes from the galactic center are compared
with the sensitivity of current telescopes (FERMI and KM3net).
The fluxes of charged antimatter cosmic rays (positrons and antiprotons) are calculated with DarkSUSY and confronted with data.
Finally, the extension of the doublet by right-handed neutrinos and consequences for neutrino 
masses and leptogenesis, are discussed in Section~\ref{sec:NeutrinLepto}.
Conclusions are drawn in Section~\ref{sec:finale}.
Appendix A contains precise discussions, for the higher multiplet cases, on the most general 
scalar potential and the differences and the similarities between complex and real multiplets. 
Appendix B gives the complete set of (co)annihilation Feynman diagrams for all the models studied in this paper.

\section{Models}
\label{sec:models}

\subsection{Inert Doublet Model}

The Inert Doublet Model (IDM) is a two Higgs doublet model with a $Z_2$ symmetry. They are
denoted by $H_1$ and $H_2$, $H_1$ being the usual Brout-Englert-Higgs doublet. All SM
particles are even under the $Z_2$ symmetry, while $H_2$ is odd. This ensures the stability
of the lightest member of $H_2$, which will be the DM candidate, and prevents from flavor
changing neutral currents (FCNC)~\cite{Deshpande:1977rw}. We will assume that $Z_2$ is not spontaneously broken,
in particular, $H_2$ does not develop a vacuum expectation value.
In order to have a neutral component, the hypercharge of a scalar doublet is necessarily
$Y=\pm 1$ (we choose to write the electric charge $Q= T_3+Y/2$). We conventionally assign
$+1$ to the hypercharge of $H_2$: 
one can write $H_2=(H^+ \quad (H_0+iA_0)/\sqrt{2})^T$,
similarly to the ordinary Higgs doublet, where $H_1=(h^+ \quad (v_0+h+iG_0)/\sqrt{2})^T$.

The most general renormalizable scalar potential with two doublets is given
by\footnote{For the doublet case, the introduction of the term
$(H_1^\dagger \tau_i H_1)(H_2^\dagger \tau_i H_2)$, where the $\tau_i$ are the $SU(2)$ generators,
 is redundant since
$$(H_1^\dagger H_1)(H_2^\dagger H_2)+(H_1^\dagger \tau_i H_1)(H_2^\dagger \tau_i H_2)
=2|H_1^\dagger H_2|^2~.$$ }

\begin{equation}\label{potential}
\begin{split}
V(H_1,H_2) &= \mu_1^2 \vert H_1\vert^2 + \mu_2^2 \vert H_2\vert^2  + \lambda_1 \vert H_1\vert^4
+ \lambda_2 \vert H_2\vert^4 \\
&+ \lambda_3 \vert H_1\vert^2 \vert H_2 \vert^2 + \lambda_4 \vert H_1^\dagger H_2\vert^2
+ {\lambda_5\over 2} \left[(H_1^\dagger H_2)^2 + h.c.\right]~.
\end{split}
\end{equation}
After the electroweak symmetry breaking, $H_1$ develops its vev, 
$v_0 = -\mu_1^2/\lambda_1 \simeq 246$ GeV, and the scalar
potential in the unitary gauge then becomes,
\bea
V &=& \frac{1}{2} m_h^2 h^2 + \lambda_1 v_0 h^3 + \frac{1}{4} \lambda_1 h^4 \nonumber \\
&& + \frac{1}{2} m_{H_0}^2 H_0^2 + \frac{1}{2} m_{A_0}^2 A_0^2 + m_{H_c}^2 H^+ H^-  \nonumber \\
&& + \frac{1}{2} \left( \lambda_{H_0} H_0^2 + \lambda_{A_0} A_0^2 + 2 \lambda_{H_c} H^+ H^- \right) \left( 2 v_0 h + h^2 \right) \nonumber \\
&& + \frac{1}{4} \lambda_2 \left( H_0^2 + A_0^2 + 2 H^+ H^-
\right)^2~, \label{brokpot} \eea with a  mass spectrum given by,
\bea
m_h^2 &=& 2 \lambda_1 v_0^2~,\cr
m_{H_0}^2 &=& \mu_2^2 +  \lambda_{H_0} v_0^2~,\cr
m_{A_0}^2 &=& \mu_2^2 +  \lambda_{A_0}  v_0^2~,\cr
m_{H^+}^2 &=& \mu_2^2 + \lambda_{H_c} v_0^2~.
\label{masses}
\eea
We have defined $\lambda_{H_c} \equiv \lambda_3/2$ and
$\lambda_{H_0,A_0} \equiv (\lambda_3 + \lambda_4 \pm
\lambda_5)/2$. We will consider $H_0$ to be the DM candidate
(i.e.~$\lambda_5<0$) though the results would be exactly the same
for $A_0$ changing the sign of $\lambda_5$.

Some theoretical constraints first apply on these quartic
couplings. To ensure that the potential is bounded from below, the
vacuum stability (at tree-level) requires that,
\bea
\lambda_{1,2} &>& 0 \quad , \nonumber \\
\lambda_{H0}~, \quad \lambda_{A_0} \, , \quad \lambda_{H_c} &>& - \sqrt{\lambda_1\lambda_2}~.
\label{vacstab}
\eea
Other constraints are also imposed from past accelerator
measurements. Indeed the extended scalar sector could bring
corrections to electroweak precision test observables (EWPT). In
particular, the variable $T$, which is a measure of the radiative corrections to $m_W/(m_Z \cos \theta_W)$, has
been calculated for the IDM~\cite{Barbieri:2006dq}
\be
\label{eq:deltaT}
\Delta T\approx \frac{1}{12\pi^{2}\alpha v^{2}}(m_{H^+}-m_{A_0})(m_{H^+}-m_{H_0})~.
\ee
However, in the high mass regime considered in this work, mass splittings turn out
to be small (see Section~\ref{sec:relic}), so that this constraint is not
limiting. We find $\Delta T \leq 6.0 \cdot 10^{-3}$, well below current
experimental bounds~\cite{Peskin:1991sw,Yao:2006px}.

The IDM has already been extensively studied in the literature. It has been shown
that a viable DM candidate with the correct relic abundance can be obtained in
three regimes, low-mass ($m_{H_0} \ll m_W$) \cite{Hambye:2007vf,Andreas:2008xy},
middle-mass ($m_{H_0} \simeq m_W$)\cite{Barbieri:2006dq,LopezHonorez:2006gr} and
high-mass ($m_{H_0} \gg m_W$)\cite{Cirelli:2005uq,LopezHonorez:2006gr}.
Direct and indirect detection constraints were investigated in
Refs.~\cite{Barbieri:2006dq,Majumdar:2006nt,LopezHonorez:2006gr,Gustafsson:2007pc,
Andreas:2008xy,Agrawal:2008xz,Andreas:2009hj,Nezri:2009jd} and confrontation of the IDM in the
$m_{H_0}<m_W$ regime to colliders data and future prospects was done in~\cite{Cao:2007rm,Lundstrom:2008ai}.
In this paper, we provide a more detailed analysis of the high mass regime and
show that the scalar coupling contribution can easily dominate over the gauge one and 
without fine-tuning (as results in Ref.~\cite{LopezHonorez:2006gr} suggested).

\subsection{Higher Multiplet Models}
\label{sec:highermultipletmodel}


The procedure followed for the doublet above can be generalized for a multiplet of
higher dimension. Let $H_n$ denotes this scalar multiplet, with $n$ being the
dimension of its representation under $SU(2)_L$. The relevant lagrangian for any
of these objects coupled to the usual Higgs doublet $H_1$ can be written as
\be
\label{eq:lagrangianmultiplet}
{\cal L} = \left(D_\mu H_n \right)^\dagger \left(D^\mu H_n\right)-V(H_n,H_1) \quad ,
\ee
with the covariant derivative given by
\be
D_\mu \equiv \partial_\mu -i g \tau^{(n)}_a W_\mu^a-i g_Y \frac{Y}{2} B_\mu \quad .
\ee
$\tau_a^{(n)}$ stands for $SU(2)_L$ generators in the
representation $\mathbf{n}$ and $Y$ is the hypercharge of $H_n$.
For $H_n$ to contain a neutral component $H_n^{(0)}$, the hypercharge $Y$ has to be odd (even)
when $n$ is even (odd).
The most general renormalizable potential for $H_n$ is given by
\begin{equation}
\label{eq:potentialmultiplet}
\begin{split}
V(H_n,H_1) = V_1(H_1)+ \mu^2 H_n^\dagger H_n &+
\frac{\lambda_2}{2} \left(H_n^\dagger H_n\right)^2 + \lambda_3
\left(H_1^\dagger H_1\right) \left(H_n^\dagger
H_n\right) \\
&+ \frac{\lambda_4}{2} \left(H_n^\dagger \tau_a^{(n)} H_n\right)^2
+ \lambda_5 \left(H_1^\dagger \tau_a^{(2)} H_1\right)
\left(H_n^\dagger \tau_a^{(n)} H_n\right)~,
\end{split}
\end{equation}
where a sum over $a$ is implicit in the last two terms. In this
potential, $\lambda_{2,3}$ are equivalent to $\lambda_{2,3}$ in
the doublet case while the $\lambda_5$ operator is equivalent to
the sum of the $\lambda_4$ and $\lambda_3$ operators in the
doublet case. The $\lambda_4$ operator reduces to the $\lambda_2$
operator in the doublet case. It is important to notice that the
$\lambda_5$ operator in the doublet case, which is responsible for
the mass splitting between $H_0$ and $A_0$, has no equivalent in
higher multiplet dimension.

Unlike the doublet case, the potential of Eq.~(\ref{eq:potentialmultiplet})
cannot give rise to a mass splitting between the real and the imaginary part of the neutral 
component of the multiplet, when this field is complex.
IF $Y\neq 0$ this would lead to a DM candidate with unsuppressed vector interactions with the $Z$ boson,
which is ruled out by direct detection limits.
Unless some mechanism is advocated to create this mass splitting,
this restricts the viable models to odd dimension multiplets with $Y=0$
(the coupling to $Z$ is proportional to $(T_3-Q \sin^2 \theta_W)$, with $Q=T_3+Y/2$).
Notice that there are still two cases with $Y=0$, depending on whether the multiplet is real or complex.
The perturbativity of $SU(2)_L$ up to the Planck scale imposes $n\leq
8$~\cite{Cirelli:2005uq}. Therefore, the only possibilities are $n=3,~5,~7$.

As in the doublet case, for $n=3$ and $n=5$, a $Z_2$ symmetry is
necessary to ensure the stability of the DM candidate. In the case
$n=7$, this parity is unnecessary because the candidate is
automatically stable. Indeed, no renormalizable or dimension 5
operator can be constructed to induce its decay into SM particles~\cite{Cirelli:2005uq}.
Moreover, an operator of dimension 6 or higher would induce a
lifetime of the order of the age of the universe or larger if the
cutoff scale is set to the GUT scale. The DM candidate for $n=7$
is accidentally stable, like the proton in the SM.\\

Let us first analyze the case of the real multiplet models.
 In a suitable basis, as detailed in Appendix~\ref{sec:annexGener}, a real multiplet
$H_n$ is written as
\be
H_n =\frac{1}{\sqrt{2}}\begin{pmatrix} \Delta^{(j_n)}
\\ \dots \\ \Delta^{(0)}\\ \dots \\
\Delta^{(-j_n)}
\end{pmatrix} \quad ,
\ee
where $j_n=(n-1)/2$, $Q$ in $\Delta^{(Q)}$ corresponds to the
electric charge, and $\Delta^{(-Q)}=\left(\Delta^{(Q)}\right)^*$.
For real multiplets, the expression $\left(H_n^\dagger \tau_a^{(n)} H_n\right)$ is identically zero. 
Therefore the terms with coefficient $\lambda_4$ and $\lambda_5$ disappear from the potential Eq.~(\ref{eq:potentialmultiplet}).
As a consequence, there is only one scalar quartic coupling ($\lambda_3$)
connecting $H_n$ to $H_1$.

After the electroweak phase transition, the SM Higgs field
develops its vev, $\langle
H_1\rangle=v_0/\sqrt2$,  and
the scalar potential in the unitary gauge becomes
\bea
V &=& \frac{1}{2} m_h^2 h^2 + \lambda_1 v_0 h^3 + \frac{1}{4} \lambda_1 h^4 \nonumber \\
&& + \frac{1}{2} \, m_0^2 \; \Delta^{(0)\,2} + \sum_{0< Q \leq j_n} m_0^2 \; \Delta^{(Q)} \Delta^{(-Q)} \nonumber \\
&& + \frac{\lambda_3}{2} \left( \frac{1}{2}\Delta^{(0)\,2} + \sum_{0< Q \leq j_n} \Delta^{(Q)} \Delta^{(-Q)} \right)( 2 v_0 h + h^2 ) \nonumber \\
&& + \frac{\lambda_2}{8} \left( \frac{1}{2}\Delta^{(0)\,2} +
\sum_{0< Q \leq j_n} \Delta^{(Q)} \Delta^{(-Q)} \right)^2~.
\label{brokpot2} \eea
At tree-level, all the multiplet components have the same mass
\be
\label{eq:massmultiplet} m^2_0 = \mu^2 + \frac{\lambda_3 v_0^2}{2}
\quad . 
\ee
At one-loop however, a mass splitting is generated by the
coupling to gauge bosons and the charged components become
slightly heavier than the neutral one~\cite{Cirelli:2005uq},
\be m\left({\Delta^{(Q)}}\right)-m\left({\Delta^{(0)}}\right) =
Q^2 \Delta M_g \quad , \label{cirsplit} \ee
where
\be \Delta M_g = g M_W \sin^2\frac{\theta_W}{2} \simeq (166 \pm
1)~\mathrm{MeV} \quad. \ee
Notice that the scalar couplings of the potential do not modify
these splittings, because they are identical for all the charged
and the neutral components. As these one-loop splittings are small, it is a
good approximation to consider all DM states as degenerate.
Finally, the vacuum stability is ensured by the condition
\bea
\lambda_{1,2} &>& 0 \quad , \nonumber \\
\lambda_3 &>& - \sqrt{2 \lambda_1\lambda_2} \quad . 
\label{vacstab2}
\eea
\\

The case of complex multiplets is analyzed in details in Appendix~\ref{sec:complexmultiplet}.
It appears that the associated phenomenology is close to the real case. 
Without the introduction of some new symmetry $U(1)$ under which $H_n$ is charged,
a complex multiplet can be decomposed into two interacting real multiplets. 

In the presence of such a symmetry, all the degrees of freedom are degenerate
at tree level except for  the $\lambda_5$ term of
Eq.~(\ref{eq:potentialmultiplet}) which induces an extra mass splitting (see
Eq.~(\ref{newmrel})) and lowers the mass of half of the charged components of
$H_n$. The neutral DM field  stays the lightest only if $\lambda_5 \lesssim 2.2 \cdot 10^{-2}$.
In the latter case, the model is similar to a real multiplet model, except for the doubling of the number 
of fields. This, in turn, reduces  the threshold mass imposed by the relic
density constraint by a factor $\sqrt{2}$ which implies that scalar DM
candidates lighter than the real case analyzed in what follows are still
allowed.

\section{Relic abundance in the high-mass regime}
\label{sec:relic}

In this section, we show that the $SU(2)_L$ scalar multiplet extension of the SM can naturally lead to a multi-TeV
DM candidate with the correct relic density.
In this high-mass regime, coannihilations play a significant role.
We will therefore start by briefly reviewing the formalism used to calculate the relic density.

\subsection{Freeze-out equations}
\label{subsec:freezeout}

To calculate the DM relic abundance, we solve the Boltzmann equation for the total
density of all the coannihilating species $n = \sum_i n_i$ (we take $i=0$ for
the lightest DM candidate
and $i>0$ for the other species),
\be
\frac{dn}{dt} + 3H n = -\langle\sigma_{{\it eff}} v\rangle (n^2-n^{{\rm eq}\;2}) \quad ,
\ee
where the effective thermal cross-section, given by
\be
\langle\sigma_{{\it eff}} v\rangle \; = \; \sum_{i,j} \langle\sigma^{ij} v\rangle \frac{n_i^{{\rm eq}}}{n^{{\rm eq}}} \frac{n_j^{{\rm eq}}}{n^{{\rm eq}}}
\label{sigmaeff}
\ee
is an average of the various thermal (co)annihilation cross-sections $\langle\sigma^{ij} v\rangle$, weighted by equilibrium densities
\be
n_i^{{\rm eq}} = \left( m_i T / 2\pi \right)^{3/2} e^{-m_i/T}   \quad .
\ee

Although a full integration of this Boltzmann equation is needed in the general case to compute the relic density,
an instructive and reliable estimate is derived from the so-called instantaneous freeze-out approximation,
when poles or thresholds don't appear in the cross-sections~\cite{Griest:1990kh}.
For cold DM, we can develop the cross-sections in the non-relativistic limit, if $\sigma v = A + B v^2$,
the corresponding thermal average is given by~\cite{Srednicki:1988ce,Gondolo:1990dk}
\be
\langle\sigma v\rangle \; \equiv \; a + b \, \langle v^2\rangle \; = \; A + 6 \;\left(B-\frac{A}{4}\right)\frac{1}{x} \quad ,
\ee
with $x=m_0/T$.
Then, the relic density is simply obtained as
\be
\Omega_{\rm DM}h^2 \simeq \frac{1.07\;10^9\;{\rm GeV}^{-1}}{J(x_F)g_*^{1/2}m_{\rm Pl}} \quad .
\label{freezout}
\ee
The post freeze-out annihilation integral $J$ is given by
\be
J(x_F)=\int_{x_F}^\infty \frac{\langle\sigma_{{\it eff}} v\rangle}{x^2}dx \quad ,
\label{Jint}
\ee
and the freeze-out point $x_F$ is found by solving the equation
\be
x_F = \ln \frac{0.0038 \; m_{\rm Pl}\; g_{{\it eff}} \;m_0 \langle\sigma_{{\it eff}} v\rangle}{(g_* x_F)^{1/2}} \quad ,
\label{xfout}
\ee
where $g_{{\it eff}} = \sum_i n_i^{{\rm eq}}/ n_0^{{\rm eq}}$ is the effective number of degrees of freedom. Usually, $x_F \simeq 25$.
When all the DM species are degenerate, the equilibrium densities for all the states are equal, and
the effective thermal cross-section $\langle\sigma_{{\it eff}} v\rangle$ is
simply the average of $\langle\sigma^{ij} v\rangle$ over all (co)annihilation channels.

\subsection{Inert Doublet Model}

In a first step, we will consider the case without quartic couplings between $H_1$ and $H_2$, and derive the relic abundance
constraint on the DM mass.
Then, in a second step, we will show how the conclusions drawn in the step one are dramatically changed when these scalar couplings are present.

\subsubsection{IDM in the pure gauge limit}
\label{subsec:pglim}

When all the quartic couplings between $H_1$ and $H_2$ vanish (except $\lambda_5$ which we take tiny but non vanishing
to avoid the direct detection problem above), all states are degenerate at tree-level.
At one-loop, the neutral states remain exactly degenerate due to the Peccei-Quinn symmetry, up to the very small $\lambda_5$
contribution, and a splitting $m_{H^+}-m_{H_0}\simeq 350$~MeV is induced
between the charged and the neutral states\cite{Cirelli:2005uq}.
Because of the smallness of these splittings for the annihilation cross-section, it is a
very good approximation to consider all states as exactly degenerate and all quartic couplings as
vanishing. In this limit, the DM species (co)annihilate into either known gauge bosons or fermions through an intermediate gauge boson.
The corresponding Feynman diagrams are shown in Fig.~\ref{GDfig} and Fig.~\ref{FDfig} of the Appendix \ref{sec:app:diagIDM}.
The only free parameter is the DM mass, so that there is a one-to-one correspondence between the DM mass and the relic density.

Let $\sigma_0 \simeq A_0/v+B_0 \, v$ be the effective cross-section in this limit, it is given by the average of the cross-sections
for all the annihilation and coannihilation processes relevant for the relic density calculation.
We obtain, at leading order,
%
\be
A_0 = \frac{(3-2\;s_w^4)\;\pi\;\alpha_2^2}{32\; c_w^4\; m_{H_0}^2} \quad ,
\label{sigmazero}
\ee
where $\alpha_2=g^2/4 \pi$ is the weak coupling constant, $s_w \equiv \sin \theta_W$, and $c_w \equiv \cos \theta_W$.
The zero velocity term $A_0$ agrees with the result for $n=2$ of Eq.~(12) in Ref.~\cite{Cirelli:2005uq},
up to a factor one half\footnote{The extra factor $1/2$ in Ref.~\cite{Cirelli:2005uq} could be due to a convention in the definition of the thermal average.}.
The velocity dependent term is mainly due to coannihilations, its coefficient $B_0$ is of the same order of magnitude as $A_0$.
The analytical expression for $B_0$ will not be given here, but we took it into account in numerical evaluations.

It is interesting to notice that all the (co)annihilation cross-sections fall as $m_{H_0}^{-2}$, as required by unitarity constraints.
For annihilations into gauge bosons, this behavior is achieved after the cancellation of various diagrams whose
amplitudes are connected by gauge invariance.
Let us for instance examine in more details the process $H_0 H_0 \rightarrow Z Z$.
Naively, the contribution from the longitudinal modes $Z_L$ to the amplitude is enhanced by a factor $m_{H_0}^2/m_Z^2$ compared to
the contribution from the transverse modes $Z_T$. This would lead to an unacceptable behavior of the cross-section, $\sigma \sim m_{H_0}^2$.
Actually, a cancellation of the longitudinal parts occurs between the $t$ and the $u$-channels on one hand, and the point-like interaction
diagram ("$p"-$channel) on the other hand. Notice that the $t$ and the $u$-channels involve the propagator of $A_0$.
When quartic couplings vanish, all DM states are degenerate, so that this cancellation is almost exact in the sense that
the residual amplitude to longitudinal modes is given by
\be
{\cal M}_L \equiv {\cal M}(H_0 H_0 \rightarrow Z_L Z_L) \simeq \frac{g^2 m_Z^2}{4 c_w^2 m_{H_0}^2} \quad .
\label{resamp}
\ee
For the two transverse modes, there is no cancellation, their amplitude amounts to
\be
{\cal M}_{Ti} \equiv {\cal M}(H_0 H_0 \rightarrow Z_{Ti} Z_{Ti}) \simeq \frac{g^2}{2 c_w^2} \quad (i=1,2) \quad .
\label{Tamp}
\ee
In the high mass regime, ${\cal M}_L \ll {\cal M}_{Ti}$.
For {\it e.g.} $m_{H_0} \simeq 550$~GeV, ${\cal M}_L/{\cal M}_{Ti} \simeq 1.4~\%$.
As $\sigma(H_0 H_0 \rightarrow Z Z) \propto \Sigma_i |{\cal M}_{Ti}|^2+|{\cal M}_L|^2$, 
we see that the residual longitudinal amplitude is totally negligible.
Therefore, in the pure gauge limit, gauge bosons produced by the annihilations of $H_0$ are almost purely transverse.

In Fig.~\ref{Omax}, we have plotted the DM relic density as a function of mass, assuming zero quartic couplings between $H_1$ and $H_2$.
The solid and the dashed curves correspond to the instantaneous freeze-out approximation with and without velocity dependent terms in $\langle\sigma v\rangle$.
As can be seen, these terms shift down the value of $\Omega_{\rm DM}h^2$ by only $\sim 4\%$.
Also shown are more exact points from a full integration of the Boltzmann equation, obtained with the MicrOMEGAs program~\cite{Belanger:2006is}.
The latest 5-years WMAP results, combined with baryon acoustic oscillations and supernovae data yield
$\Omega_{\rm DM} h^2 = 0.1131 \pm 0.0034$~\cite{Hinshaw:2008kr}. We see that, in the absence of quartic couplings, the DM mass is determined by the relic density,
\be
m_{H_0}=534 \pm 8.5\;{\rm GeV}\;(1\sigma) \quad ,
\label{pgomega}
\ee
in agreement with the results of Ref.~\cite{Cirelli:2005uq}, up to the update of $\Omega_{\rm DM} h^2$.
At $3\sigma$ the DM mass cannot be lighter than $508$~GeV. It is worth noticing that the value of $m_{H_0}$
is quite sensitive to the precision at which $\Omega_{\rm DM}h^2$ is determined. Also, the approximate solid curve from Fig.~\ref{Omax} gives a slightly
higher mass range, $m_{H_0}=553 \pm 8.5$~GeV. The discrepancy is attributable to the instantaneous freeze-out approximation rather than to the values
of the cross-sections in Eq.~(\ref{sigmazero}).
\bfig
\bc
\includegraphics[width=0.6\textwidth]{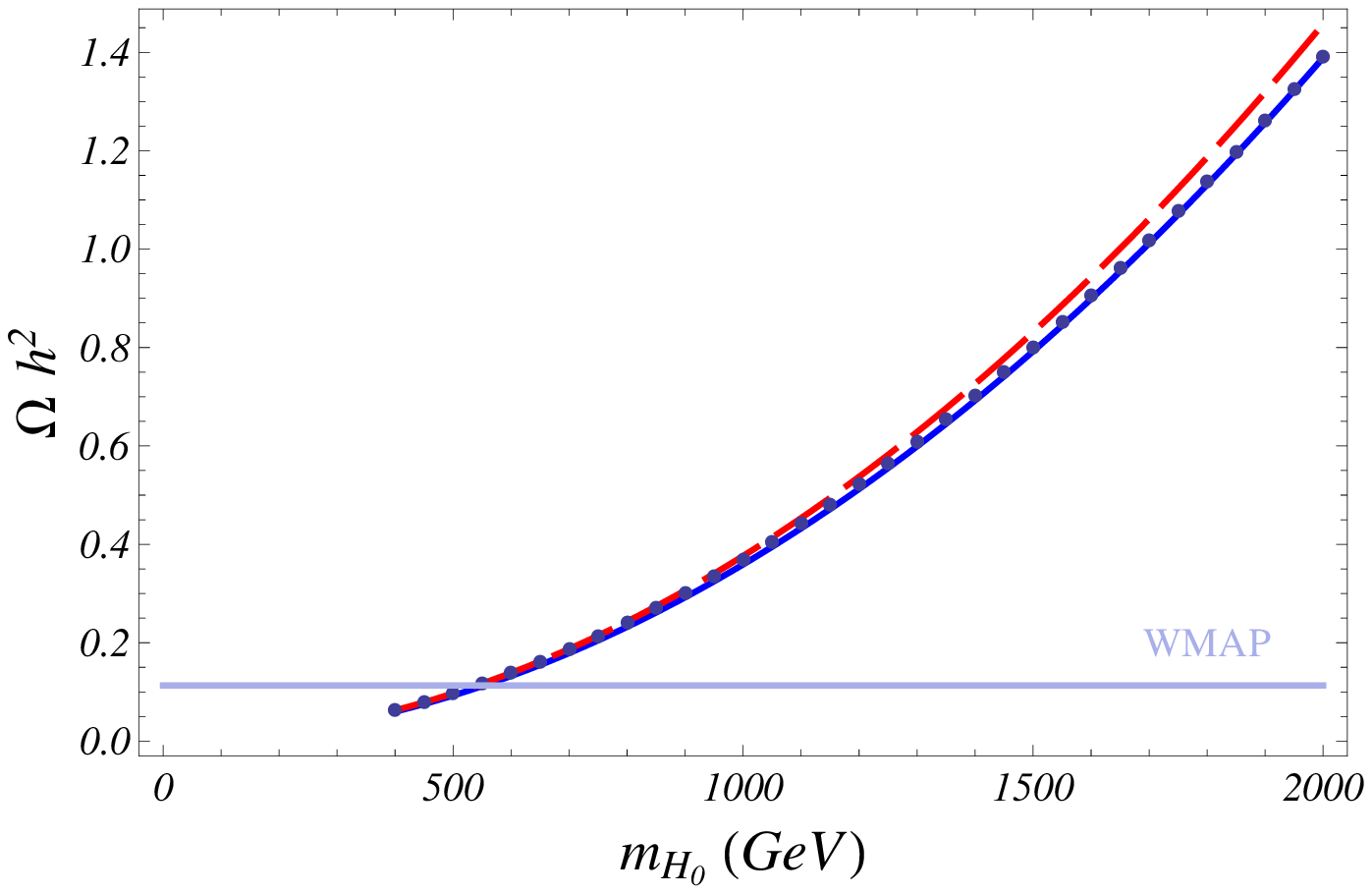}
\caption{{\it Dark matter relic abundance in the pure gauge limit as a function of the DM mass. \textcolor{Red}{Dashed} (\textcolor{Blue}{Solid}) curve~:
Instantaneous freeze-out approximation without (with) velocity-dependent terms in $\sigma v$. Points : Output from MicrOMEGAs}}
\label{Omax}
\ec
\efig

\subsubsection{Effect of the quartic couplings}
\label{subsec:quartcoupl}

When the scalar quartic couplings between $H_1$ and $H_2$ are switched on, the cross-section is affected in two ways.
First, non-zero mass splittings between members of the inert doublet, Eq.~(\ref{masses}) will modify the amplitude of pure gauge diagrams
of both annihilation and coannihilation processes.
Second, a series of  new annihilation and coannihilation processes which involve the usual Higgs particle appear, see Fig.~\ref{HDfig} of the Appendix \ref{sec:app:diagIDM}.

It is instructive to analyze how cross-sections grow with the quartic couplings.
As a generic example, let us consider again the process $H_0 H_0 \rightarrow ZZ$.
When a non zero mass splitting between $H_0$ and $A_0$ exists, the amplitudes to 
longitudinal modes from the
$t$, $u$ and $p$-channels do not cancel exactly anymore 
and is not suppressed anymore by a $m^2_Z/m^2_{H_0}$ as in Eq.~(\ref{resamp}).
Instead, a contribution proportional to $(m_{A_0}^2-m_{H_0}^2)$ remains
(in the high-mass regime squared mass splittings are small with respect to $m^2_{H^0}$ but not necessarily with respect to $m^2_Z$, as we will see),
%
\be
{\cal M}_L^\lambda \owns \frac{g^2}{2 c_w^2} \cdot \frac{m_{H_0}^2}{m_Z^2} \cdot \frac{m_{A_0}^2-m_{H_0}^2}{m_{H_0}^2} \quad .
\label{amplambda}
\ee
Furthermore, there is also a new contribution from the Higgs exchange in the $s$-channel
$H_0 H_0 \rightarrow h^* \rightarrow ZZ$.
The amplitude of the $s$-channel is proportional to $\lambda_{H_0} v_0^2 = (m_{H_0}^2-\mu_2^2)$. When added to the $t$, $u$ and $p$-channels,
a further cancellation takes places, and the total amplitude to longitudinal modes due to quartic couplings is finally given by
\bea
{\cal M}_L^\lambda &\simeq& \frac{g^2}{2 c_w^2} \cdot \frac{m_{H_0}^2}{m_Z^2} \cdot 
\frac{\left\{(m_{A_0}^2-m_{H_0}^2)+(m_{H_0}^2-\mu_2^2)\right\}}{m_{H_0}^2} \nonumber \\
&=& 2 \lambda_{A_0}
\label{amplambda2}
\eea
to be compared with Eqs.~(\ref{resamp}-\ref{Tamp}). 
Corrections to transverse modes are negligible, because they are smaller by a factor $m_Z^2/m_{H_0}^2$.
Therefore, gauge bosons produced by the scalar quartic couplings are almost purely longitudinal
in the high mass regime, whereas those from the gauge interactions alone are almost purely transverse. 
As a consequence, the annihilation cross-section can only grow (if we neglect the tiny residual amplitude
of Eq.~(\ref{resamp})) when scalar quartic couplings are switched on.
The scalar coupling contribution to the cross-section $\sigma(H_0 H_0 \rightarrow ZZ)$
becomes comparable to the gauge one for $\lambda_{A_0} \simeq g^2/(2 \sqrt{2}c_w^2) \simeq 0.2$.
This corresponds to a small value of the splitting $\vert m_{A_0}-\mu_2 \vert \sim m_W^2/\mu_2$. 

The above analysis for the process $H_0 H_0 \rightarrow ZZ$ serves 
as a demonstration that the various (co)annihilations 
cross-sections can only grow with the splittings between $\mu_2$, 
$H_0$, $A_0$ and $H_c$ ($H_c$ stands for $H_{\pm}$).
To further check this conclusion, we will make use of an expansion 
of the cross-sections that is valid in the asymptotic high-mass 
regime we consider here.
The cross-sections are simultaneously expanded in 
$m^2_{W,Z,h}/m^2_{H_0}$ and in $\lambda_{H_0,A_0,H_c} v_0^2/m_{H_0}^2$
(except maybe for the top quark, the corrections induced by the 
fermion masses of the SM are really negligible in this regime).
The orders of magnitude of these parameters which give the correct 
relic abundance will serve as an {\it a posteriori} justification 
for the use of this expansion, as we will see.
We separate the various inclusive (co)annihilation cross-sections 
into $\lambda$ independent and $\lambda$ dependent terms as
\be
\sigma^{ij} = \sigma_0^{ij} + \sigma^{ij}_\lambda \quad ,
\ee
with $\{i,j=0,1,2,3\}$ corresponding to $\{H_0, A_0, H^+, H^-\}$.
Only the expressions of the dominant velocity independent terms will be given here.
To leading order, for $\sigma_0^{ij}$, we have
\bea
A_0^{11} = A_0^{22} = A_0^{34} &=& \frac{(1+2 c_w^4)\,g^4}{128\pi c_w^4 m_{H_0}^2}~, \nonumber \\
A_0^{13} = A_0^{14} = A_0^{23} = A_0^{24} &=& \frac{s_w^2 g^4}{64\pi c_w^2 m_{H_0}^2}~.
\eea
The $\lambda$ dependent cross-sections can be written in a compact way as
\be
\sigma^{ij}_\lambda \equiv \frac{\Lambda^{ij}}{32\pi m_{H_0}^2} \quad ,
\ee
with the coefficients $\Lambda^{ij}=\Lambda^{ji}$ given by
\bea
\Lambda^{00} = \Lambda^{11} &=& 2(\lambda_{H_0}^2 + \lambda_{A_0}^2 + 2 \lambda_{H_c}^2) \nonumber \\
\Lambda^{22} = \Lambda^{33} = 2\Lambda^{01} &=& 2(\lambda_{H_0} - \lambda_{A_0})^2 \nonumber \\
\Lambda^{02} = \Lambda^{03} = \Lambda^{12} = \Lambda^{13} &=& (\lambda_{H_0} - \lambda_{H_c})^2 + (\lambda_{A_0} - \lambda_{H_c})^2 \nonumber \\
\Lambda^{23} &=& (\lambda_{H_0} + \lambda_{A_0})^2 + 4 \lambda_{H_c}^2 \quad .
\label{sigmalambda}
\eea
As we can see, the cross-sections $\sigma^{ij}$ can only increase when the scalar quartic couplings are switched on.
The values of the $\lambda$ corresponding to a constant cross-section $\sigma^{00}$ (or $\sigma^{11}$) lie on an ellipsoid.
For the others $\sigma^{ij}$, the ellipsoid is degenerate in a cone ($\Lambda^{02}$ or $\Lambda^{23}$), or in a plane ($\Lambda^{22}$).
We can notice in Eq.~(\ref{sigmalambda}) that the coannihilations cross-sections $\sigma_\lambda^{01}$, $\sigma_\lambda^{02}$
and $\sigma_\lambda^{12}$ are determined by the mass splittings between $H_0$, $A_0$ and $H_c$,
while the annihilation cross-sections $\sigma_\lambda^{00}$, $\sigma_\lambda^{11}$ and $\sigma_\lambda^{23}$
also depend on the splitting between $H_0$, $A_0$, $H_c$ and the scale $\mu_2$.

From the positivity of the coefficients $\Lambda^{ij}$ in Eq.~(\ref{sigmalambda}), we can expect that the relic density will decrease
when the quartic couplings are turned on.
As shown by the instantaneous freeze-out approximation, the final relic abundance is actually controlled by the
effective thermal cross-section Eq.~(\ref{sigmaeff}), where Boltzmann suppression factors $e^{-(m_i-m_0)/T}$ appear when the mass splittings differ from zero.
The net result between this thermal damping effect and the rise of each cross-section with $\lambda$ turns out to be positive.
Therefore, even in the presence of non zero scalar couplings, the lower bound
$m_{H_0} \geq m^*$ for the relic density, with
\be
m^* = 534 \pm 25 \; {\rm GeV} \; (3\sigma) \quad ,
\ee
remains valid. Above this threshold, the scalar coupling contribution to the cross-section has to become progressively 
dominant over the gauge one in order to obtain the correct relic density set by WMAP. 
It can be seen from Fig.~\ref{Omax} or Eq.~(\ref{sigmazero}) that both contributions become equal for 
$m_{H_0} \simeq \sqrt{2} m^* = 755$~GeV. For $m_{H_0} \geq 1.7$~TeV, the gauge contribution falls below 10\%.  
\bfig
\bc
\bt{cc}
\includegraphics[width=0.45\textwidth]{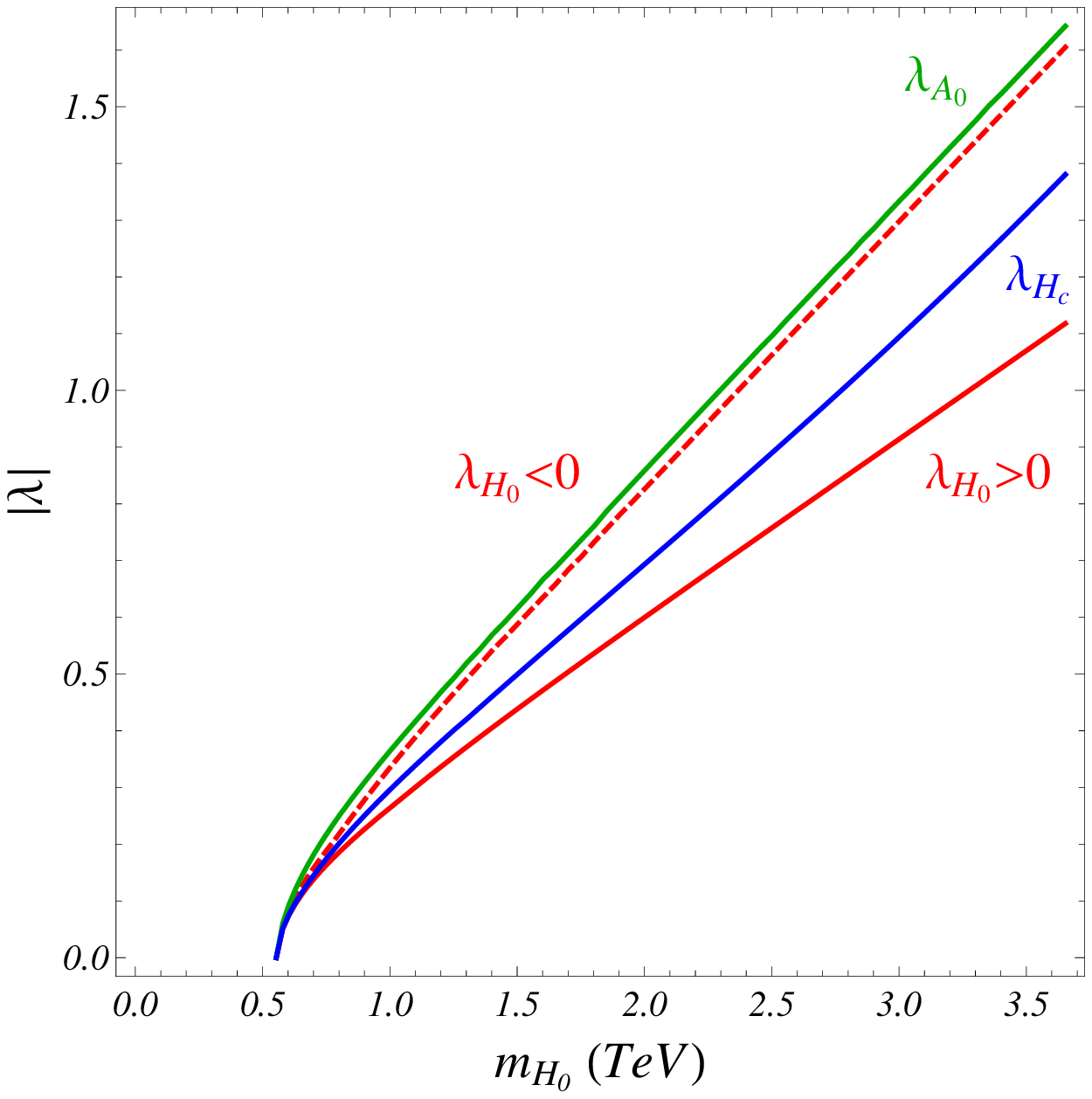}  \quad &
\includegraphics[width=0.45\textwidth]{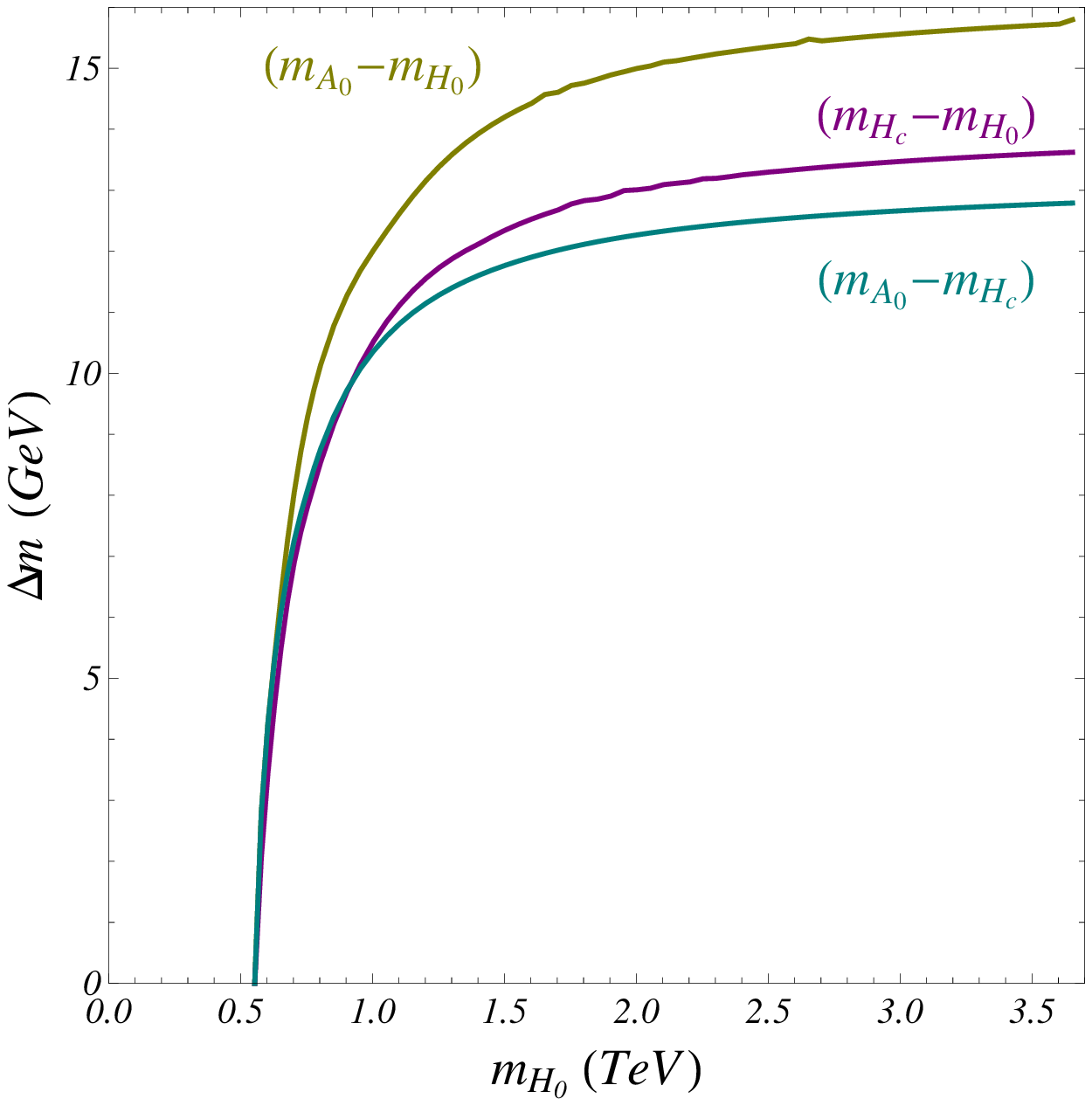}  \quad
\et
\caption{{\it Maximal scalar quartic couplings (left panel) and mass splittings (right panel) as a function of mass, imposed by the WMAP bound.
Notice that $|\lambda_{H_0}|$ is maximal for negative values of $\lambda_{H_0}$.
Asymptotic values of the maximum splittings are given in Eq.~(\ref{maxsplitval}).}}
\label{lambdamax}
\ec
\efig

For a given mass $m_{H_0}$, it is clear that the values of the quartic couplings that are compatible with WMAP are bounded.
As we will see, due to Eq.~(\ref{sigmalambda}), they form approximately an ellipsoid in the parameter space $\{\lambda_{H_0},\lambda_{A_0},\lambda_{H_c}\}$.
Mass splittings are also limited, because
$$\Delta m_{ij} \equiv m_i-m_j \simeq (m_i^2-m_j^2)/2\mu_2 \propto (\lambda_i - \lambda_j)~.$$
Upper bounds for each $|\lambda|$ and for each mass splitting are shown in Fig.~\ref{lambdamax}, as a function of $m_{H_0}$.
We see that large values of $\lambda \simeq {\cal O}(1)$ are permitted, the upper bound for each $\lambda$ (or a linear combination
of them) grows linearly with the DM mass when $m_{H_0} \gg m^*$.
We have noticed earlier that the coannihilation cross-sections are determined by the mass splittings.
It turns out that coannihilations involving the charged component are stronger for a given mass splitting.
This explains why the maximum mass splittings between $H_c$ and $H_0$ (or $A_0$) are slightly smaller than the maximum splitting $m_{A_0}-m_{H_0}$.
Also the maximum $m_{H_c}-m_{H_0}$ is larger than the maximum $m_{A_0}-m_{H_c}$ because we have assumed that $H_0$ is the lightest DM particle.
As $\Delta m_{ij} \simeq (m_i^2-m_j^2)/2\mu_2\simeq (\lambda_i - \lambda_j)v_0^2/2m_{H_0}$ and $\lambda_i - \lambda_j \propto m_{H_0}$
for $m_{H_0} \gg m^*$, each mass splitting is bounded by an asymptotic value.
Numerically, we find
\bea
|m_{A_0}-m_{H_0}| &<& 16.9\;{\rm GeV} \quad , \nonumber \\
|m_{H_c}-m_{H_0}| &<& 14.6\;{\rm GeV} \quad , \nonumber \\
|m_{A_0}-m_{H_c}| &<& 13.6\;{\rm GeV} \quad .
\label{maxsplitval}
\eea
These small splittings serve as an a posteriori justification of the pertinence of the joint expansion in
$m^2_{W,Z,h,f}/m^2_{H_0}$ and in $\lambda_{H_0,A_0,H_c} v_0^2/m_{H_0}^2$ we made. They also imply that corrections
to EWPT observables are negligible in the high mass regime of the IDM (see Eq.~(\ref{eq:deltaT})).

On Fig.~\ref{lambdacontours}, two sections in the parameter region allowed by WMAP are shown for three values of the DM mass
$m_{H_0}=600,\,1000,\,3000$~GeV, they correspond to the two cases $m_{A_0} = m_{H_0}$ and $m_{H_c} = (m_{A_0}+m_{H_0})/2$.
The contours were obtained using the instantaneous freeze-out approximation and for a $1\sigma$ variation of $\Omega_{\rm DM} h^2$.
We have checked that the results agree very well with the output of MicrOMEGAs.
Ellipsoidal contours are superimposed on these figures in red dashed line.
They correspond to an expansion in terms of $\Delta m_{ij}/T$ up to quadratic terms of the Boltzmann exponential factors in the effective thermal
cross-section. This ellipsoidal approximation is not accurate when the mass splittings are
not negligible compared to the temperature around freeze-out, as can be seen for example in the right panel of
Fig.~\ref{lambdacontours} when $m_{H_0}=600$~GeV.
\bfig
\bc
\bt{cc}
\includegraphics[width=0.45\textwidth]{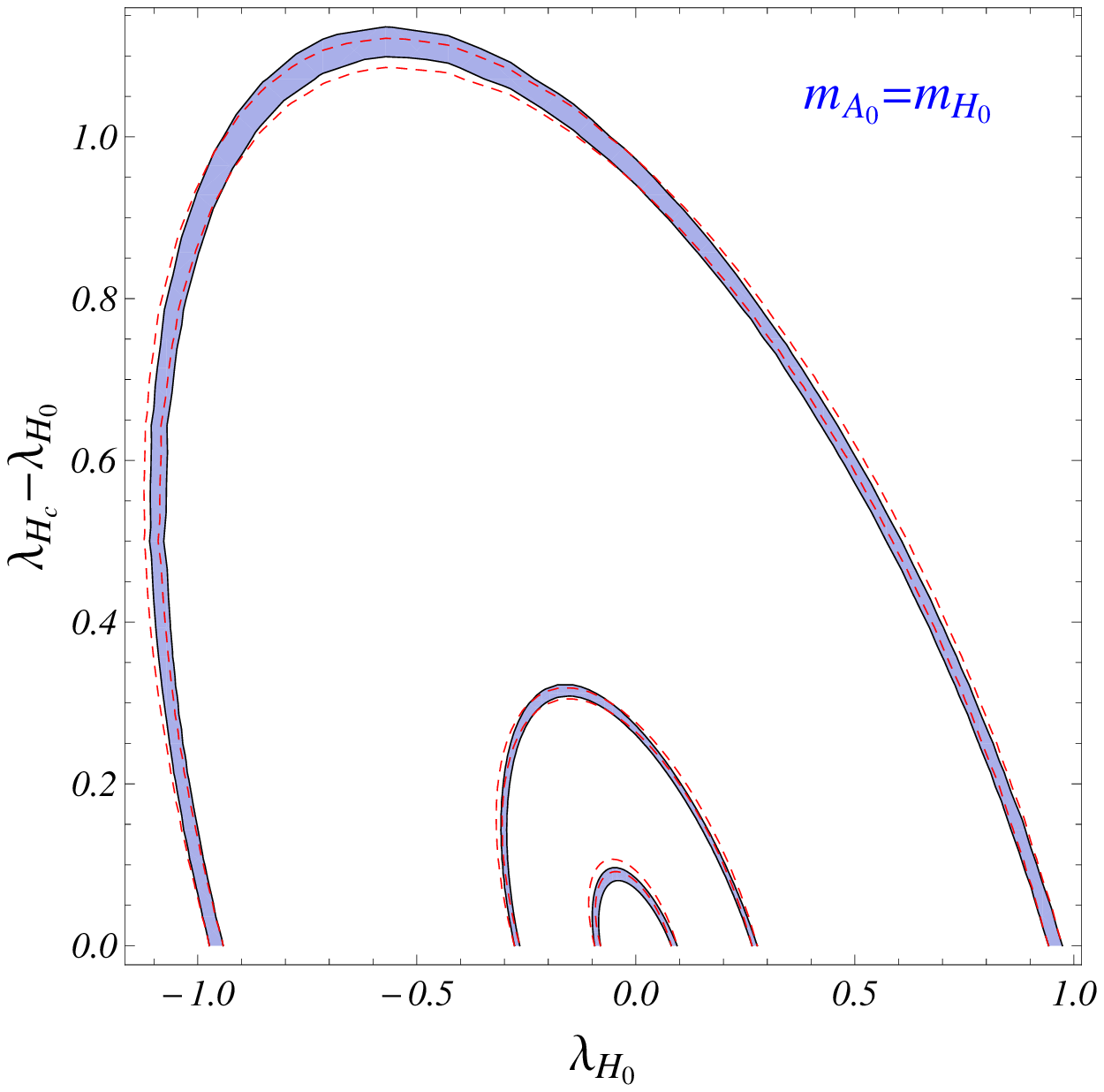}  \quad &
\includegraphics[width=0.45\textwidth]{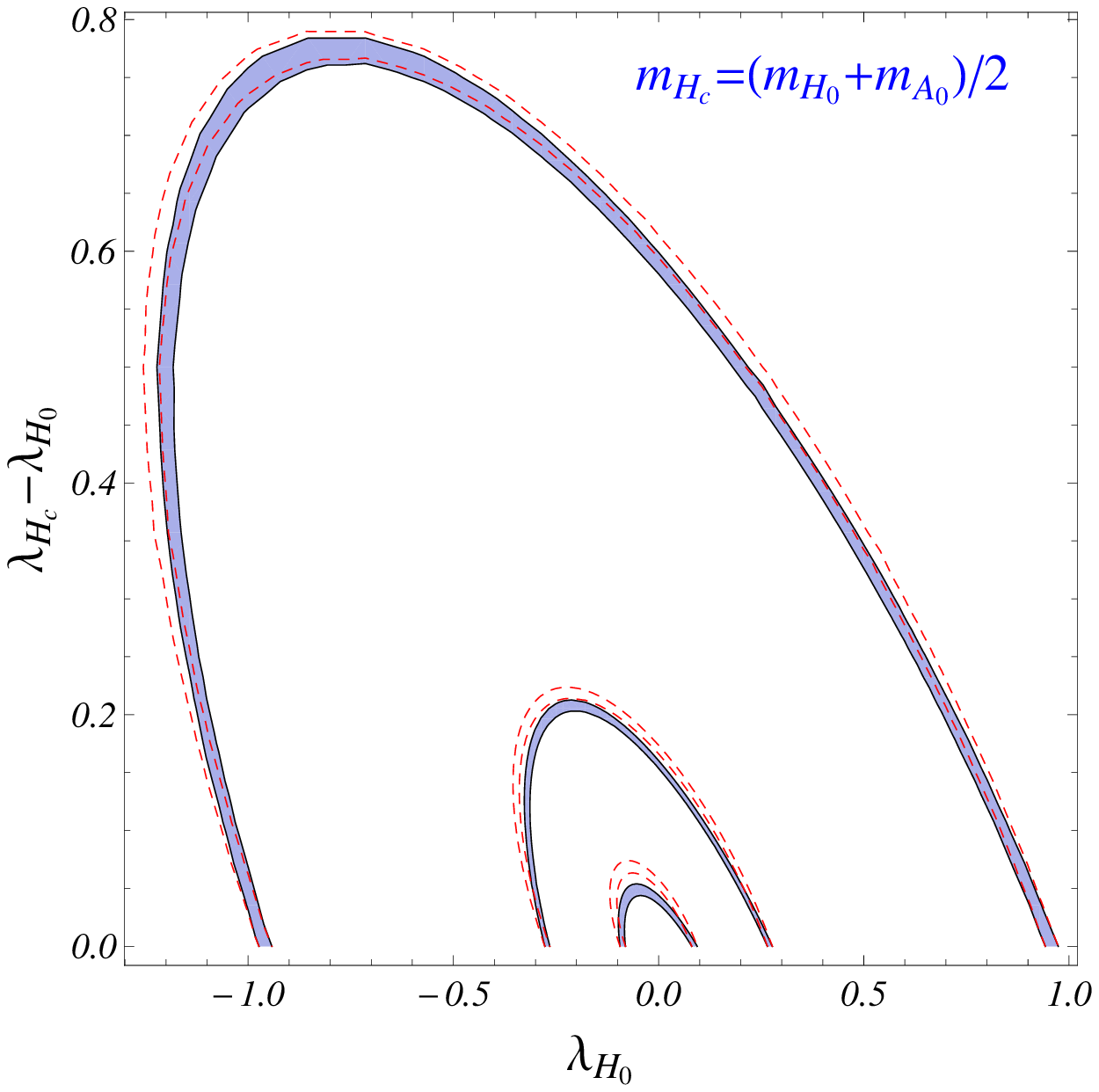}  \quad
\et
\caption{{\it Contours of $\lambda$ for the WMAP value $\Omega_{\rm DM} h^2 = 0.1131
  \pm 0.0034$ for $m_{H_0}=600$ (interior), $1000,\;3000$ (exterior)
  GeV,  
with $m_{A_0}=m_{H_0}$ (left panel) and $m_{H_c}=(m_{H_0}+m_{A_0})/2$ (right panel). 
\textcolor{Red}{Dashed} curve corresponds to the approximate ellipsoid.}}
\label{lambdacontours}
\ec
\efig

We can conclude that the relic abundance required by WMAP can be naturally achieved in the large mass regime of the inert doublet model.
A viable DM candidate with a mass in the multi-TeV range only requires at least one of the scalar quartic couplings to be of order~1.
In this case, there is no need for any fine-tuning in the parameters of the lagrangian.
Moreover, the relic density constraint does not put a lower bound on any of the quartic couplings (or a linear combination of them).
For any value of the DM mass above the limit of Eq.~(\ref{pgomega}), one or two of them can even be zero accidentally or if some symmetry is added to the model.

The growth of the values of $\lambda$ with mass as needed by WMAP imposes an upper bound on the mass of the DM candidate because
of unitarity considerations.
If we require all the physical quartic couplings $\lambda_{H_0}$, $\lambda_{A_0}$ and $\lambda_{H_c}$ in Eq.~(\ref{brokpot}) to be smaller than $4\pi$,
we get the bound
\be
m_{H_0} <  58\;{\rm TeV} \quad ,
\ee
while for $2\pi$ instead of $4\pi$ we get $m_{H_0}<30$~TeV.
This is in agreement with the general unitarity bound which holds on any thermal DM relic whose relic density proceed
from the freeze-out of its annihilation \cite{Griest:1989wd}.

\subsubsection{Freeze-out during the unbroken phase of the Standard Model}

We have shown that the inert doublet model in the high mass regime can provide a viable DM candidate for a very
large range of the DM mass. 
However, for large masses above $\sim 5$~TeV, the freeze-out will occur before the onset of the electroweak phase transition.
In this case, all DM components have the same mass $m_0=\mu_2$ at the epoch of freeze-out, 
so that deviations from the previous analysis are expected.
They annihilate into components of the usual Higgs doublet $H_1$ in the unbroken phase.
The threshold between the unbroken and the broken phases occurs at a temperature $T_c \simeq 200$~GeV. 
Although the electroweak phase transition in the SM is second order the phase transition, it occurs rather quickly (see e.g.~\cite{Burnier:2005hp} and Refs. therein) and for simplicity
we will assume a sharp threshold for the freeze-out of our DM candidate.
This means that the freeze-out will be assumed to be in the broken (unbroken) phase for $m_{H_0} \leq (>) 5$~TeV.   

In the unbroken phase, the scalar coupling part of the effective annihilation cross-section relevant 
for the relic density is modified 
as\footnote{We assume that all the components of the Higgs doublet have masses much smaller than $m_0$.}
\be
\sigma_\lambda = \frac{\lambda_3^2+\lambda_4^2+\lambda_5^2}{64 \pi m_0^2} \equiv \frac{r_\lambda^2}{16\pi m_0^2} \quad ,
\ee
the pure gauge part of the cross-section is still given by Eq.~(\ref{sigmazero}), in the limit $s_w \rightarrow 0$.
The WMAP constraint determines $r_\lambda$, we get
\be
r_\lambda \simeq 2.85 \left( \frac{m_0}{10~TeV} \right)
\ee
Note that $r_\lambda=4\pi$ for $m_0=46$~TeV and $r_\lambda=2\pi$ for $m_0=22.5$~TeV.
The DM mass range is therefore slightly reduced.

The maximal values of the scalar quartic couplings and the mass splittings allowed by WMAP corresponding to a freeze-out in the unbroken phase
are shown in Fig.~\ref{highlambdamax}. We see that $\vert \lambda_{H_0} \vert$ and $\lambda_{A_0}$ are increased while $\lambda_{H_c}$ is reduced.
However, if the stability conditions are taken into account, the maximum of $\vert \lambda_{H_0} \vert$ is given by the positive branch $\lambda_{H_0}>0$
which is smaller. Also, the maximum mass splitting $\vert m_{H_0}-m_{A_0} \vert$ can be slightly higher if the freeze-out occurs during the unbroken phase of the SM.
We get
\be
|m_{A_0}-m_{H_0}| < 17.6\;{\rm GeV} \quad .
\ee
Finally, as it can be seen on Fig.~\ref{highlambdamax}, the vaccum stability conditions affect significantly the maximal values of
$|m_{A_0}-m_{H_0}|$ and $|m_{H_c}-m_{H_0}|$.
%
%
%
%
\bfig
\bc
\bt{cc}
\includegraphics[width=0.45\textwidth]{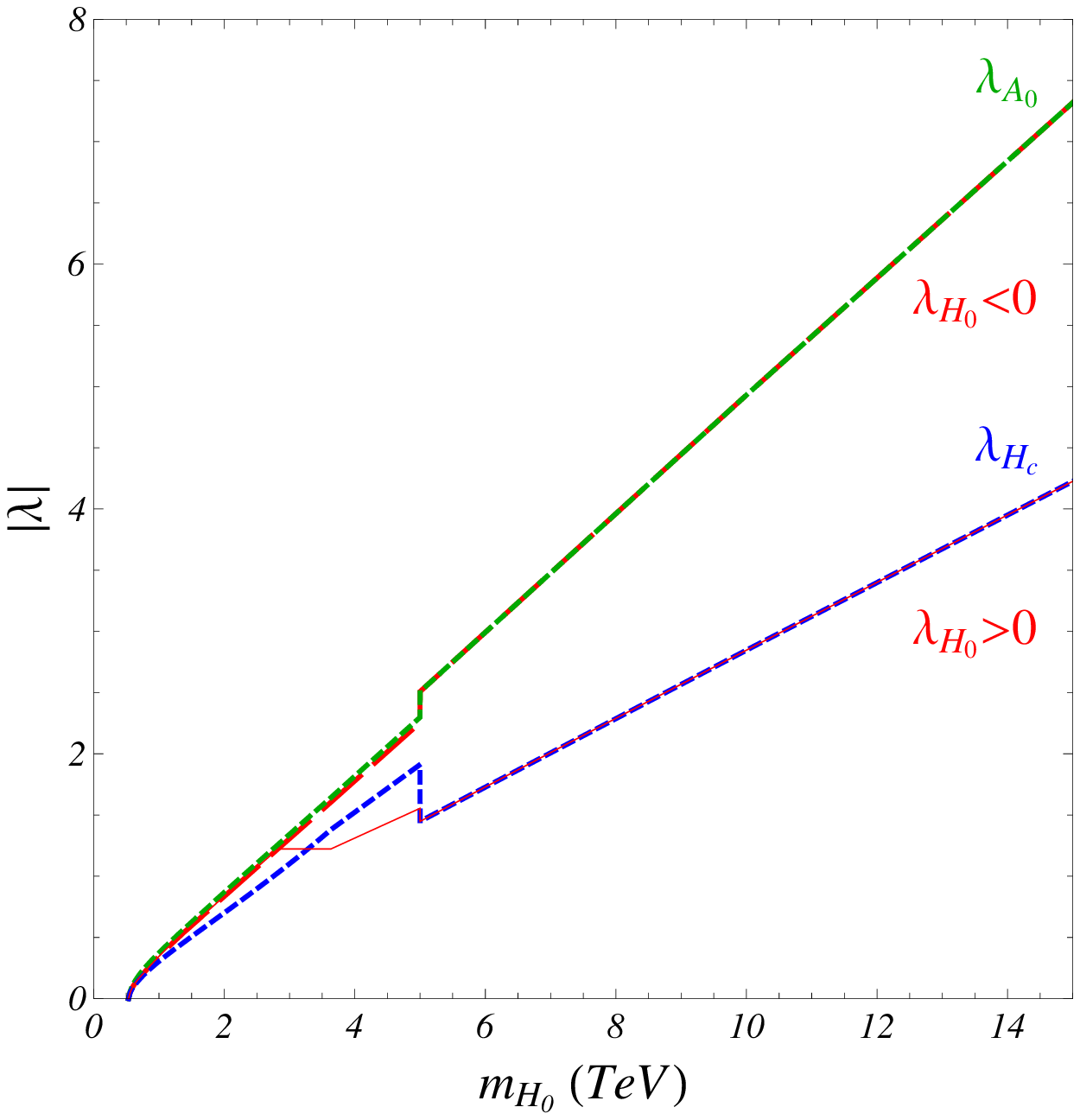}  \quad &
\includegraphics[width=0.45\textwidth]{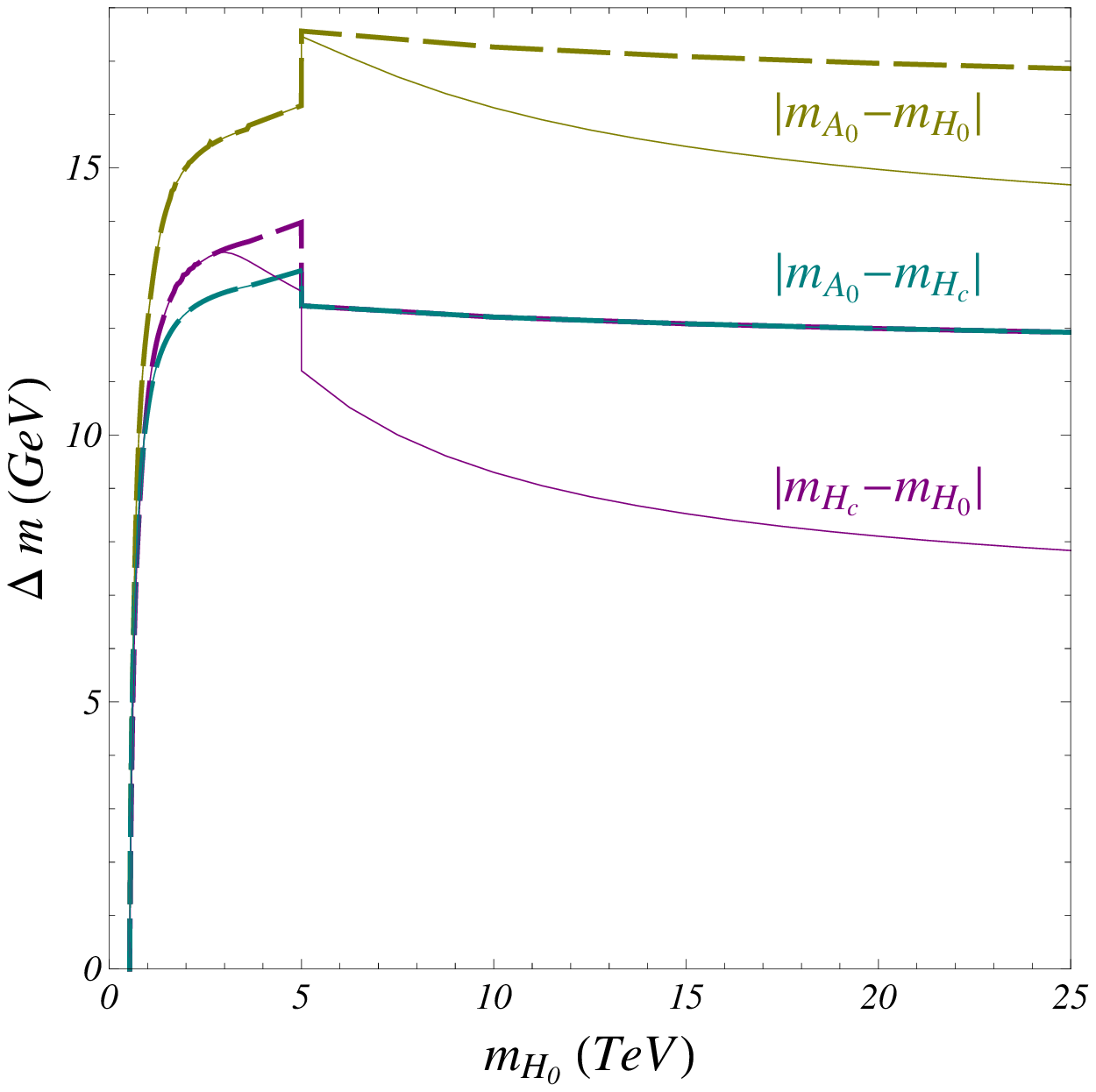}  \quad
\et
\caption{{\it Maximal values of scalar quartic couplings (left panel) and mass splittings (right panel) as a function of the DM mass, constrained by WMAP,
without (dashed lines) and with (thin solid lines) the vacuum stability conditions Eq.~(\ref{vacstab}) included.
We assume a Higgs mass $m_h=120$~GeV, and a sharp threshold between the freeze-out in the broken and in the unbroken phases of the SM
at a mass $m_{H_0}=5$~TeV.}}
\label{highlambdamax}
\ec
\efig
It is worth emphasizing that the stability conditions do not constrain the mass range of the DM candidate.
To fulfill them, it suffices for $\lambda_3$ to be positive
and larger in absolute value than $\lambda_4+\lambda_5$. Therefore, these conditions do not put a stringent constraint
on the possibility of having a very heavy DM candidate with the correct relic abundance.

\subsection{Higher multiplet case}\label{sec:HMRD}
\subsubsection{(Co)Annihilation cross-sections}

In the following we consider only the case of real multiplets. As explained in section \ref{sec:highermultipletmodel}, for complex multiplets one just must  divide the mass obtained with a real multiplet by $\sqrt{2}$. 
At tree-level, all the components of a real multiplet have the same mass $m_0$.
As the mass splittings induced at one-loop are very small (see Eq.~(\ref{cirsplit})),
it is a very good approximation to consider all states as exactly degenerate.
The effective cross-section used for the calculation of the relic density
is therefore the average of all annihilation and coannihilation cross-sections between the odd particles
composing the multiplet. The Feynman diagrams of all the relevant processes are depicted
in the figures of Appendix~\ref{sec:Feynmangraphmultiplet}.

For higher multiplets, the only scalar quartic coupling that has an influence on the relic density is $\lambda_3$.
As for the inert doublet, we can develop the cross-sections in a simultaneous expansion in
$m^2_{W,Z,h,f}/m^2_0$ and in $\lambda_3 v_0^2/m_0^2$.
At leading order, the dominant velocity independent terms of the effective annihilation
cross-section $\sigma^{(n)}v = \sigma^{(n)}_0v + \sigma^{(n)}_\lambda v$
are\footnote{Like in the doublet case, the expression of $A_0^{(n)}$ in Eq.~(\ref{eq:crossectanalyt1}) differs from
the result of Ref.~\cite{Cirelli:2005uq} by a factor $1/2$.}
\be
\label{eq:crossectanalyt1}
A_0^{(n)} = \frac{(n^2-1)(n^2-3)}{n}\, \frac{g^4}{128 \pi\; m_0^2} \qquad
\mathrm{and} \qquad A_\lambda^{(n)} = \frac{1}{n} \frac{\lambda_3^2}{16 \pi\; m_0^2} \quad.
\ee
They drop as $m_0^{-2}$, as expected from unitarity considerations.
In the pure gauge limit $\lambda_3 = 0$, the odd DM components annihilate almost exclusively into gauge bosons.
Coannihilations into fermion final states are $p$-wave suppressed.
When $\lambda_3 \neq 0$, new channels of (co)annihilation are opened, through a Higgs particle, or into Higgs particles.
Expressions for the velocity dependent terms in the cross-section will not be given, as they are subdominant.
We have checked numerically that they lead to a maximal correction smaller than about $5\%$.
They have been taken into account in a numerical evaluation  of the relic density with the instantaneous
freeze-out approximation.

Notice that for high multiplets ($n>2$), the  high mass regime we consider
($m_{DM}>m_W$) is the only possibility for a successful DM
phenomenology. 
Below $m_W$, given the collider bounds on the charged multiplet
component and given the small neutral-charged component mass splittings,
coannihilation cross sections would be far too large to account for WMAP DM
abundance.

\subsubsection{Relic density}

The relic density in both real and complex models with $n=3,5,7$ has been computed using MicrOMEGAs~\cite{Belanger:2006is},
and compared to the result of the instantaneous freeze-out approximation.
The agreement between the two approaches is better than $4.5\,\%$.
For a real multiplet of a given dimension $n$, the relic abundance $\Omega_{\rm DM} h^2$ depends
only on the two free parameters of the model, the mass of the DM candidate $m_0$ and the coupling $\lambda_3$.
Therefore the WMAP constraint on the relic density determines $\lambda_3$
as a function of $m_0$, or vice-versa.
The values of $m_0$ corresponding to $\lambda_3 = 0,\,2\pi,\,4\pi$ are given in Table~\ref{tab:HighMultip}.
We find threshold masses ({\it i.e.} for $\lambda_3 = 0$) that are systematically smaller than the values obtained
in Ref.~\cite{Cirelli:2005uq} by $\sim 10\%$.
\begin{table}
\small
\bc
\bt{lccccc}
\hline \hline
Models & $\lambda_3=0 $ & $\lambda_3=2\pi $ & $\lambda_3=4\pi $ & $\lambda_3=0$ (SE) & $\lambda_3=4\pi$ (SE) \\
\hline
Real Triplet & $1.826 \pm 0.028$ & $11.1$ & $21.9$ & 2.3 & 28.1\\
Real Quintuplet & $4.642 \pm 0.072$ & $9.6$ & $17.4$ & 9.4 & 35.7\\
Real Septuplet & $7.935 \pm 0.12$ & $10.6$ & $16.1$ & 22.4 & 46.3\\
\hline \hline
\et
\ec
\caption{{\it Threshold masses (in TeV) without or with Sommerfeld effect (SE) for higher multiplet models, as 
determined by the WMAP constraint, the errors quoted correspond to a 1$\sigma$ variation of the relic density.
The large mass range of the DM candidate is shown by the indicative values for $\lambda_3=2\pi$ and $4\pi$.}}
\label{tab:HighMultip}
\end{table}

The values of the parameters $m_0$ and $\lambda_3$ that are in agreement with the WMAP
constraint at $1\,\sigma$ level are shown on Fig.~\ref{fig:M0delambda3} for
all real multiplet models. Similarly to the doublet case,
the following mass-coupling relations hold: $(m_0-m_0^*) \propto \lambda_3^2$ for
$m_0$ close to the threshold value $m_0^*$, while $\lambda_3 \propto m_0$ for $m_0 \gg m_0^*$.
This behaviour is easily recovered from the analytic expression of the effective cross-section, Eq.~(\ref{eq:crossectanalyt1}),
because the WMAP constraint fixes its value.
More precisely, we see that for $m_0 \gg m_0^*$, the DM mass scales like $m_0 \sim \lambda_3/\sqrt{n}$,
which explains the different slopes of the linear part of the function $m_0(\lambda_3)$ (see Fig.~\ref{fig:M0delambda3}).
\bfig
\bc
\includegraphics[width=0.45\textwidth]{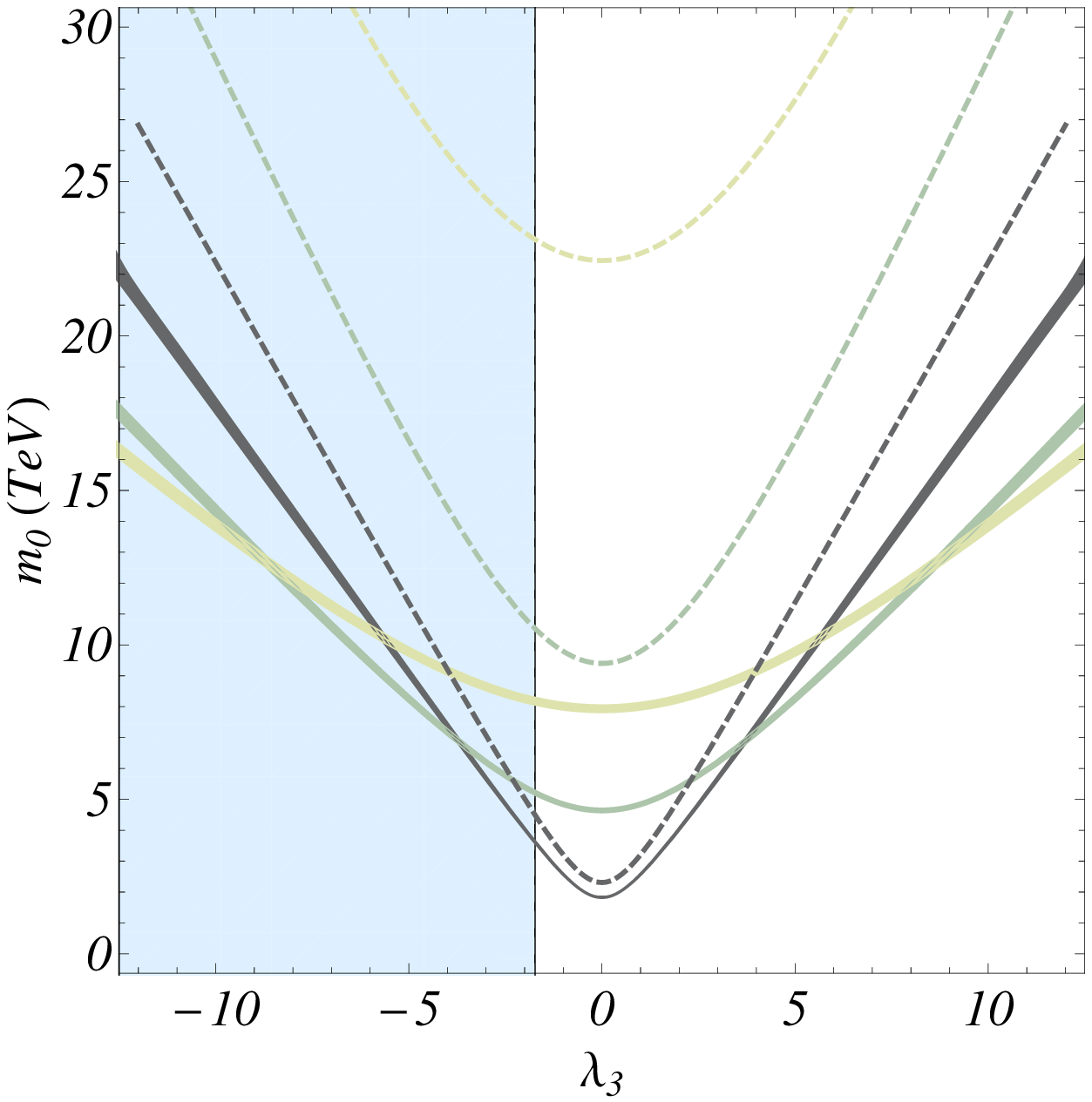}
\caption{{\it Evolution of the mass of the dark matter candidate as a function of the
coupling $\lambda_3$ for all the higher multiplet models of phenomenological interest,
as constrained by WMAP, without (solid lines) or with (dashed lines) Sommerfeld effect. 
The curves correspond, from top to bottom at $\lambda_3 = 0$, to the real septuplet, the
real quintuplet and the real triplet. The shaded area on the left is excluded by the vacuum stability constraint
(for $m_h=120$~GeV and $\lambda_2^{max}=4\pi$).}}
\label{fig:M0delambda3}
\ec
\efig

An upper bound on $m_{\rm DM}$ can be obtained by demanding that the
theory stays perturbative. Values of $m_0$ corresponding to $\lambda_3 = 2 \pi$
and $\lambda_3 = 4 \pi$ are given in Table~\ref{tab:HighMultip}. 
%
%
For higher multiplet models, the allowed DM mass is in the multi-TeV range, even when the Sommerfeld corrections are not included.
For all candidates with a mass higher than around 5 TeV, the freeze-out will occur in the unbroken phase of the SM.
Unlike the doublet case, the expressions for the effective cross-sections
given by Eqs.~(\ref{eq:crossectanalyt1}) for the broken phase remain valid in the 
unbroken one, although the detailed (co)annihilation processes are different.
Therefore, the behaviour of $m_0$ as a function of $\lambda_3$ given in Fig.~\ref{fig:M0delambda3} is still valid.
For higher multiplets, the so-called Sommerfeld effect plays however a significant role. 

\subsubsection{Sommerfeld effect}

At small relative velocity, the interaction between two particles becomes long range if the mass of the particle exchanged in the interaction 
is much smaller than the two interacting particles masses. 
This leads to a non perturbative enhancement of the annihilation cross-sections of very
heavy DM candidates, known as the Sommerfeld effect~\cite{Sommerfeld:1931}.
For an abelian vector interaction, the two parameters which determine the strength of the enhancement are 
$\alpha/\epsilon \equiv m_{DM}/(m_V/\alpha)$ and $\alpha/\beta$,
where $m_V$ is the mass of the vector particle, $\alpha$ is the coupling constant and $\beta$ is the DM velocity.
For non abelian interactions, the general trend of the enhancement is controlled by the same parameters than in the abelian case, but
is complicated by the possibility of resonances~\cite{Hisano:2004ds}.
It has been shown in Ref.~\cite{Cirelli:2007xd} that Sommerfeld effect corrections affect both the relic density calculations and the present day annihilation
cross-sections relevant for indirect detection signals.

A full treatment of the Sommerfeld corrections is beyond the scope of the present paper. In the case of heavy scalar dark matter candidates,
corrections to both weak and scalar interactions are \emph{a priori} expected. 
However, given the fact that the scalar interactions give contributions either pointlike or, from the exchange of a Higgs boson, suppressed by a $v^2_0/m_{DM}^2$ factor,
in what follows, we will only take into account the Sommerfeld corrections due to gauge bosons,
which has been calculated in details in Ref.~\cite{Cirelli:2007xd}. 
As their full computation shows, away from resonances, the
enhancement factor at the time of freeze-out is almost constant. Moreover, this factor strongly depends on the multiplet dimension.
For the doublet, it is negligible (a few percent), but it increases rapidly for higher multiplets.
The first result can be understood by the fact that thermal contributions to the gauge boson masses were included. 
If $m_V \propto g T$, the parameter $\alpha/\epsilon$ is roughly constant at the time of freeze-out $T \simeq m_{DM}/x_F$.
The second result can be understood by the fact that (co)annihilation cross-sections grow roughly as $n^2$ with the dimension of the multiplet
(see Eq.~(\ref{sigmaDI}) for the annihilation of $\Delta_0$). Effectively, the coupling constant is therefore enhanced by the size
of the multiplet. Numerically, we find that $n^2-1=8,~24,~48$ for $n=3,~5,~7$
scales well as the enhancement factor $\simeq 1.6,~4.1,~8.0$.   

In Fig.~\ref{fig:M0delambda3}, we show how the Sommerfeld effect modifies the relation between $m_0$ and $\lambda_3$ for all higher multiplets.
Following the line of reasoning of the last paragraph, a simple approximation has been used. A constant cross-section enhancement factor
is taken for each multiplet. It is chosen so as to reproduce the threshold masses given in Ref.~\cite{Cirelli:2007xd} 
(see Table~\ref{tab:HighMultip}) when Sommerfeld corrections are taken into account. 
Resonances as well as the possible enhancement factor from scalar couplings have been neglected in this 
simple treatment. Such effects would push the DM mass to even higher values for a given value of $\lambda_3$.

\section{Direct detection} \label{sec:DD}

The neutral scalar field of an $SU(2)_L$ multiplet has vector-like interactions with the Z boson if its hypercharge $Y \neq 0$.
This leads to an elastic spin independent cross-section between the DM candidate
and the nucleon that is 2 to 3 orders of magnitude above current limits~\cite{Ahmed:2008eu,Alner:2007ja}.
To evade this constraint, either the multiplet has $Y=0$, or a mass splitting (at least of the order of $\simeq 100$~keV)
is induced by some mechanism between the real and the imaginary components of the neutral field.
In the latter case, the DM - nucleon interaction through the Z boson is kinematically forbidden, or leads to tiny inelastic collisions.
The doublet case is special in this respect, because the most general renormalisable potential Eq.~(\ref{potential}) automatically allows for such
mass splitting. This direct detection constraint explains why we only considered the inert doublet model and higher multiplets with $n=3,5,7$ in this work.

For the models studied, the only tree-level interaction between the DM candidate and the nucleon proceeds through the exchange of a Higgs scalar,
which gives rise to elastic spin independent collisions. If $\lambda_h$ is the coupling between the DM and the Higgs particle,
the DM-nucleon cross-section is then given by
\be
\sigma_{\rm DM \, N}^\lambda \simeq f_N^2 \frac{\lambda_h^2}{\pi} \left( \frac{m_N^2}{m_{\rm DM} m_h^2} \right)^2 \quad ,
\label{sigmadd}
\ee
where $f_N \simeq 0.3$ is the nucleonic form factor determined
experimentally (see~\cite{Ellis:2000ds} and also, for a
range for $f_N$, see~\cite{Andreas:2008xy} and references therein.).
For the doublet case, $\lambda_h \equiv \lambda_{H_0}$ while $\lambda_h \equiv \lambda_3/2$ for higher multiplets.

Elastic scatterings through the exchange of $SU(2)_L \times U(1)_Y$ gauge bosons is also possible, but only at loop level.
Its contribution to the elastic cross-section has been computed in Ref.~\cite{Cirelli:2005uq},
\be
\sigma_{\rm DM \, N}^0 = f_N^2(n^2-1)^2 \frac{\pi \alpha_2^4 m_N^4}{64 m_W^2}\left(\frac{1}{m_W^2}+\frac{1}{m_h^2}\right)^2
\label{sigmaDDgauge}
\ee
The value of this pure gauge part increases rapidly with the dimension $n$ of the multiplet, and does not depend on the DM mass
for a given multiplet.

\subsection{Doublet}

The mass splitting between $H_0$ and $A_0$ is controlled by $\lambda_5$.
As we have seen, the relic density constraint does not put a lower bound on the absolute value of this parameter.
Tiny mass splittings are allowed, and are stable against radiative corrections.
This opens the interesting possibility of explaining the DAMA annual modulation data while still being compatible with other
experimental bounds through inelastic collisions with the nucleon $H_0 \; n \rightarrow A_0 \; n$~\cite{Fu2009}.
It has been shown that the mass splitting $|m_{A_0}-m_{H_0}|$ should be roughly in the range $[50...150]$~KeV to realize this scenario (see e.g.~\cite{Cui:2009xq} and Refs.~therein).
For a DM mass between 535~GeV and 10~TeV, this corresponds to a tiny value of $\lambda_5$,
\be
0.9 \cdot 10^{-6} \leq \lambda_5 \leq 5 \cdot 10^{-5} \quad ,
\ee
a range that also has important possible implications on leptogenesis and neutrino masses (see Section~\ref{sec:NeutrinLepto} below).
A precise determination of the parameter range that fits all direct detection data strongly depends on assumptions on the velocity
distribution of the dark matter particles in the Earth neighborhood, and will not be presented here~\cite{Fu2009}.
In any case, the inelastic collisions rapidly become inoperant when the mass splitting is increased above 1~MeV.

In the case where the inelastic collisions through a Z boson can be neglected, we are left 
with the elastic cross-sections of Eqs.~(\ref{sigmadd}-\ref{sigmaDDgauge}).
The maximum value of $\lambda_{H_0}$ allowed by the relic density constraint grows linearly with $m_{H_0}$,
this translates into an absolute upper bound on the elastic cross-section.
Numerically, we find
\be
\sigma_{H_0 \, N} < 9.4 \cdot 10^{-9} \; {\rm pb} \quad .
\ee
\bfig[t]
\bc
\bt{cc}
\includegraphics[width=0.45\textwidth]{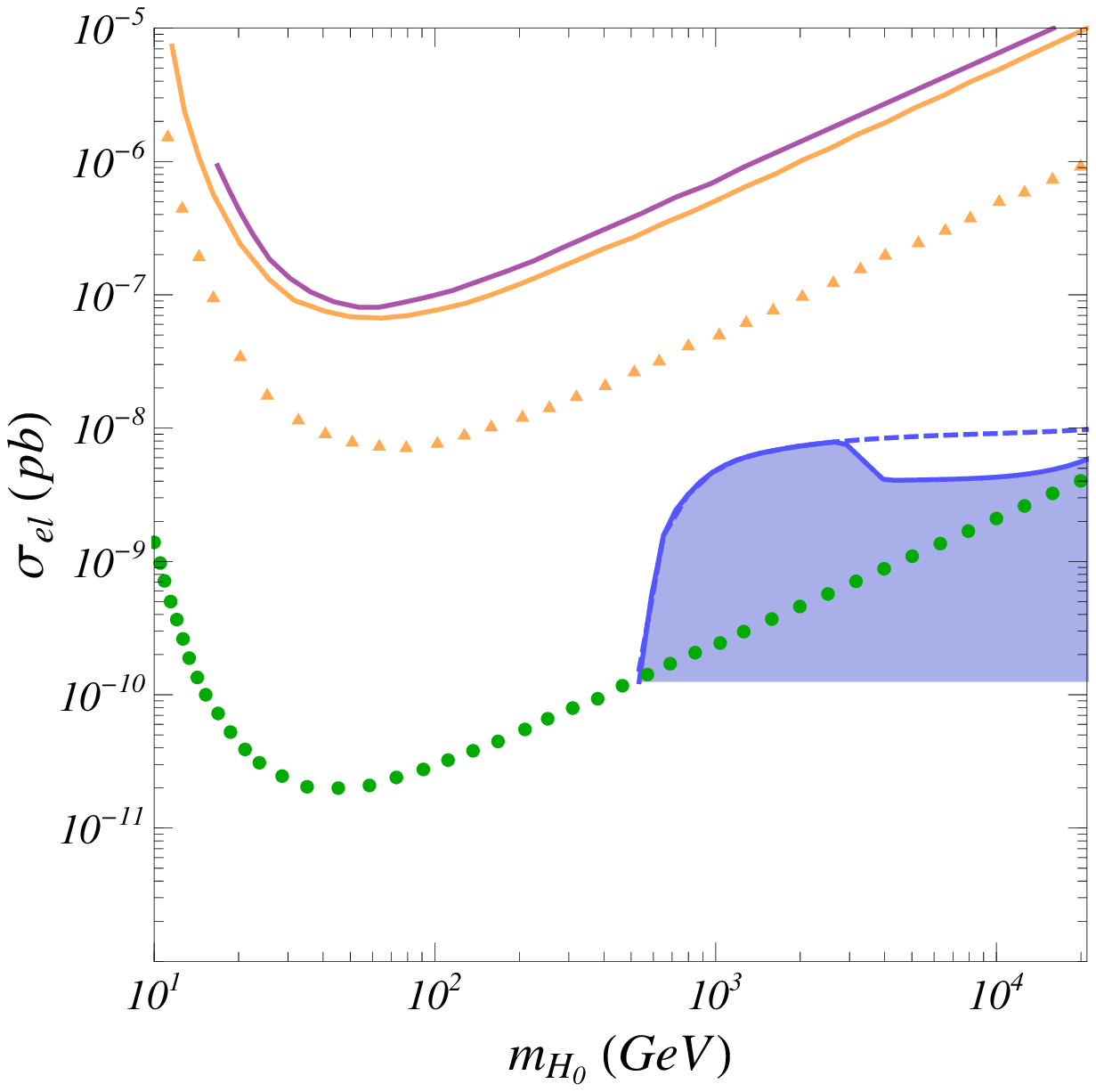} \quad &
\includegraphics[width=0.45\textwidth]{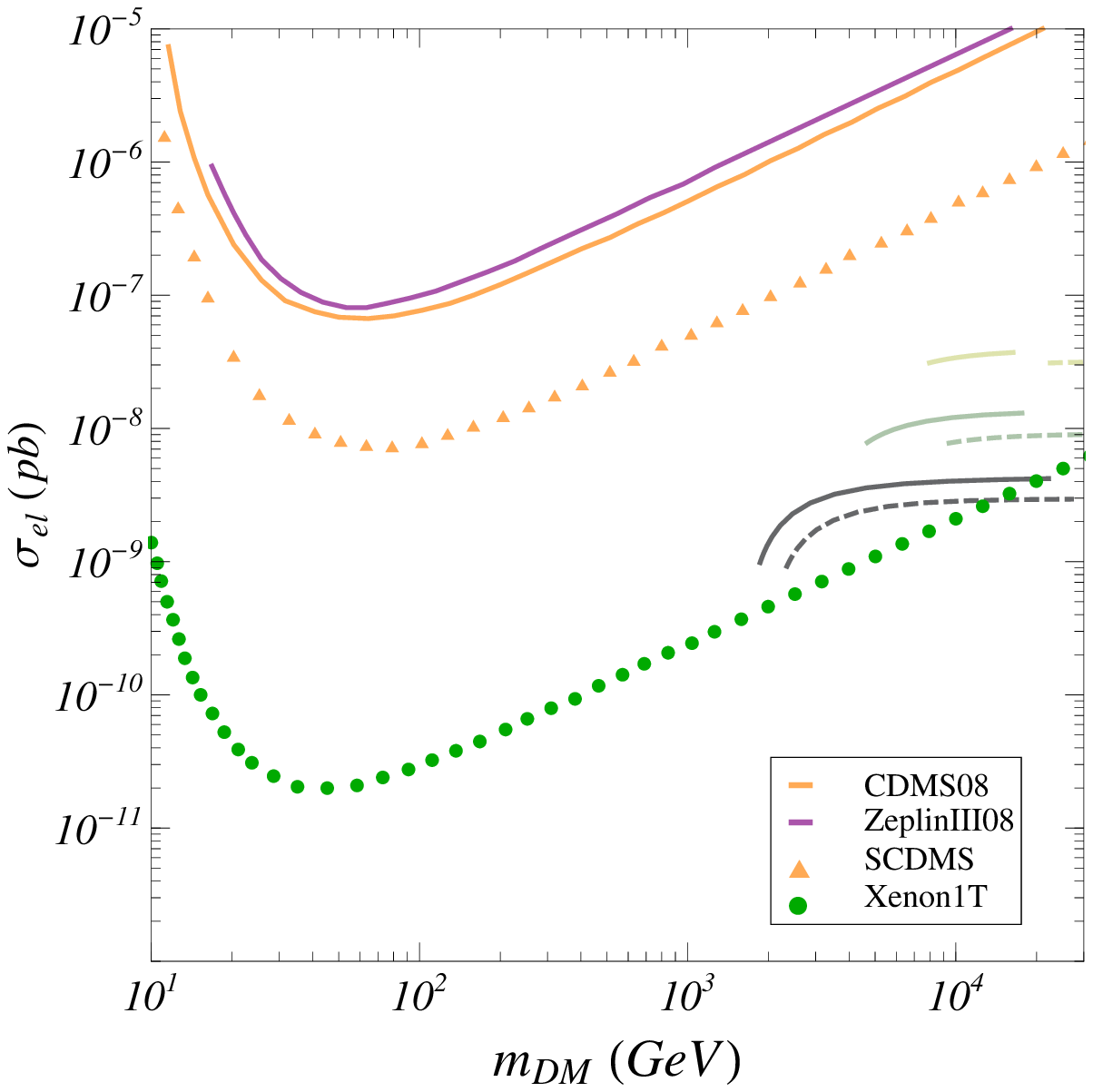} \quad
\et
\caption{{\it Elastic cross-section on nucleon for the inert doublet (left panel) and for higher multiplets (right panel),
compared to experimental limits (CDMS Ge result from 2008~\cite{Ahmed:2008eu},
Zeplin III final result (2008)~\cite{Alner:2007ja}) and projected sensitivities at future experiments 
(Super-CDMS and Xenon 1T)~\cite{DMtools}.
We have assumed $m_h=120$~GeV, a standard Maxwellian DM halo with a local density $\rho_0=0.3~{\rm GeV/cm^3}$.
For the left panel the shaded area gives the allowed range of values. Its lower limit corresponds to the pure gauge interaction cross section putting all quartic interactions to 0. The upper limit on the elastic cross-section is given by the solid (dashed) blue line when
vaccuum stability conditions are (not) taken into account.
For the right panel, solid (dashed) curves correspond to the cross-section prediction without (with) Sommerfeld effects.}}
\label{dddmax}
\ec
\efig
While on one hand an upper bound is derived from the WMAP constraint, on the other hand, the pure gauge cross-section Eq.~(\ref{sigmaDDgauge})
sets a lower bound around $10^{-10}$~pb.
As a result, the direct detection rate can vary by two orders of magnitude, for a given DM density and velocity distribution around the
earth neighborhood.
In Fig.~\ref{dddmax}, we show the range of the elastic cross-section as a function of mass, compared to current experimental bounds and
future experiments sensitivity. The DM-nucleon cross-section has been calculated with $m_h=120$~GeV.
The limits and projections assume a standard value for the local DM density $\rho_0=0.3$~GeV$/{\rm cm}^3$, and a maxwellian velocity distribution
with the characteristic halo velocity $v_{halo}=220~{\rm km/s}$.
Extending the expected XENON reach to $10$~TeV, we see that a large portion of the parameter space of the IDM in the high mass regime
will be probed by future direct detection experiments with a 1~Ton $\times$ year sensitivity.

\subsection{Higher multiplet}

In the case of higher multiplets, no mass splitting between the neutral
components of a complex multiplet can be generated with the scalar
potential of Eq.~(\ref{eq:potentialmultiplet}).
As a result, only multiplets with vanishing hypercharge $Y=0$ are viable.
Therefore, we are led to consider only real multiplets of dimension $n=3,5,7$.
For a given DM mass, the only free parameter is $\lambda_3$, which is determined by the relic density constraint (see Fig.~\ref{fig:M0delambda3}).
Higher multiplet models are therefore particularly predictive.

As for the doublet case, the spin independent elastic scattering cross-section has two parts.
The pure gauge part given by Eq.~(\ref{sigmaDDgauge}) yields
\be
\sigma_{\rm DM \, N}^0 = (0.86,~7.76,~31.03) \times 10^{-9} \; {\rm pb} \quad 
\mathrm{for}\quad n=3,5,7~.
\ee
The scalar quartic coupling part increases with $m_{\rm DM}$, with the following upper bounds,
\be
\sigma_{\rm DM \, N}^\lambda < (3.34,~5.28,~6.15) \times 10^{-9} \; {\rm pb} \quad 
\mathrm{for}\quad n=3,5,7~.
\ee
The total cross-section is represented in Fig.~\ref{dddmax}. 
It is well above the sensitivity limit of the future XENON 1T experiment but below current limits.
Contrarily to the doublet case, the gauge contribution to the elastic cross-section is always substantial or even dominant
compared to the scalar one. When Sommerfeld corrections to the relic density calculation
are taken into account, the relative contribution of the scalar interactions to the
elastic cross-section is even smaller (see Fig.~\ref{dddmax}).  In all cases for a given value of $m_{DM}$ the direct detection cross section is predicted to one value.

\section{Indirect detection}
\label{sec:ID}

The annihilation of DM can produce several types of signals useful for indirect detection searches. 
We will examine photons and neutrinos, which give a directional signal, and also charged antimatter cosmic rays
for which such a directional information is lost after diffusive processes.
Heavy scalar DM particles mainly annihilate into $Z Z$, $W^+ W^-$ and $h h$,
which subsequently produce the desired signal in cascade decays.
Therefore, the annihilation of heavy scalar candidates generally produces soft spectrums.

In what follows, we will only derive predictions for these soft spectrums.
The monochromatic signal from direct annihilation into photons at one loop for example will not be considered here,
as this cross-section is strongly affected by non perturbative effects when it becomes non negligible~\cite{Hisano:2002fk}.
Generally speaking, a detailed analysis of Sommerfeld enhancements and resonance effects is beyond the scope of the present paper,
so that all predictions will be made with an enhancement factor $EF=1$. It is however worth noticing that 
heavy scalar DM models considered here are viable for a wide range of mass when scalar interactions are taken into account.
As a result, the phenomenon of resonances, which occurs for particular values of the DM mass, can always be achieved by
tuning the DM mass to one of these values.
Such a possibility does not appear for fermionic minimal dark matter candidates, where the DM mass is determined by the
relic density constraint~\cite{Cirelli:2005uq}.

We will now give a brief description of the flux calculation for each type of indirect signal. Then, the predictions for the inert doublet
and the higher multiplet models will be presented.

\subsection{$\gamma$ and $\nu$ signals}

The galactic center (GC), where the DM halo surrounding
the galactic disk is believed to be the most concentrated, is the most promising
region to probe for DM annihilations.
If the halo is very cuspy, as simulations~\cite{Navarro:1996gj,Kravtsov:1997dp,Moore:1999gc} 
and some dynamical mechanisms~\cite{Klypin:2001xu} suggest, observation of photons or neutrinos
from the GC can provide a clean signature of the presence of DM.
However, if the halo has a rather flat profile, as indicated by direct kinematical observations~\cite{Binney:2001wu},
the signal might be difficult to disentangle from the astrophysical background.

The total flux of $\gamma$ or $\nu$ in a solid 
angle $\Delta \Omega$ around the galactic center is simply calculated as
\be
\Phi_{\gamma,\nu}(\Delta \Omega) =  \frac{\langle \sigma v \rangle}
{2 m_{\rm DM}^2} \; N_{\gamma,\nu} \times
\frac{\Delta \Omega \, \rho_0^2 \, R_0}{4\pi} \; \bar{J}(\Delta \Omega) \quad ,
\label{fluxGC}
\ee
where
\be
N_{\gamma,\nu} = \int_{E_{min}}^{E_{max}}\sum_i
 \frac{dN^i_{\gamma,\nu}}{dE} BR_i~,
\ee
is the average number of $\gamma$ or $\nu$ per annihilation with 
an energy between the experimental thresholds $E_{min}$ and $E_{max}$,
$\rho_0$ is the local DM density, and
\be
\bar{J}(\Delta \Omega) = \frac{BF}{\Delta \Omega \, \rho^2_0 \, 
R_0} \int\,\rho^2\,\, d l \, d\Omega~,
\ee
is a dimensionless astrophysical factor which encodes all the 
uncertainties about the distribution of DM in the galactic halo.
The quantity $BF$ is the so-called (astrophysical) boost factor, an enhancement 
factor due to the clumpiness of DM in galactic halos.
It should however be stressed that the boost factor is dependent 
upon the observation direction.
For the direction of the galactic center, $BF$ is negligible if 
the halo is very cuspy.
For a flat profile like the isothermal one, $BF$ is also limited 
because the concentration of subhalos is comparable to
that of the galactic halo~\cite{Athanassoula:2008fn}. 
On the particle physics side, the Sommerfeld effect can provide non negligible enhancements. 
The predictions made hereafter do not include any boost.

For the sake of completeness, let us finally mention that in the case of neutrinos, 
an interesting possibility is to search for an signal from the core of the Sun or of the Earth,
emanating from annihilations of DM particles captured by the celestial body.
However, in the high mass regime, the capture rate of DM particles, 
which scales as $m_{\rm DM}^{-4}$ is too small to lead to observable 
signals. Also, in the case of the Earth, it cannot be enhanced by 
resonance effects like for lighter candidates with a mass around 
$50$~GeV~\cite{Andreas:2009hj}.

\subsection{Charged antimatter cosmic ray signals}

As antimatter cosmic rays (CR) are quite rare in the galaxy, they are also promising messengers to probe for exotic physics,
like DM annihilations. 
The recent publication of the positron fraction observed by PAMELA~\cite{Adriani:2008zr}, together with the excess
seen by the ATIC experiment~\cite{:2008zzr} have triggered a lot of activity in this research field, as they point to
a positron excess between 10 and 800 GeV~\cite{Cholis:2008wq}.
The explanation of both excesses by a DM scenario would lead to a candidate with rather unusual properties~\cite{Cirelli:2008pk},
and is therefore not favored. In this paper, typical flux spectrums for both positrons and antiprotons are presented,
but no attempt will be made to fit the PAMELA or the ATIC excess. 

The diffusive random walk of charged charged cosmic rays through the galaxy can be described by the following general (steady state) 
propagation equation~\cite{Lavalle:2008zb}:
\begin{equation}
  \vec{\nabla} \left[ K(E, \vec x)\vec{\nabla} {\cal N}_{\rm cr} -
    \vec{V}_{\rm conv} {\cal N}_{\rm cr} \right]
  +\frac{\partial}{\partial E}\left[b(E){\cal N}_{\rm cr} +
    K_{EE} \frac{\partial}{\partial E}  {\cal N}_{\rm cr} \right]
  + \Gamma(E) {\cal N}_{\rm cr} + {\cal Q}
   =  0\; 
  \label{eq:propag_crs}
\end{equation}
where ${\cal N}_{\rm cr}\equiv dn_{\rm cr}(E)/dE$ is the number density per unit of energy, $K(E, \vec x)$ and $K_{EE}$ are coefficients that encode 
the diffusion by galactic magnetic fields in real and momentum space, $\vec{V}_{\rm conv}$ is the velocity of the galactic convective wind,
$b(E)$ is the rate of energy loss, $\Gamma(E)$ accounts for spallation processes (destruction of CR due to collisions with the interstellar medium), and 
${\cal Q}$ is the source term. For positrons, the dominant processes are the energy loss and spatial diffusions. For antiprotons,
the dominant processes beside diffusion are spallation and convection. For a detailed discussion of the propagation model,
we refer the reader to Ref.~\cite{Moskalenko:1997gh,Baltz:1998xv,Delahaye:2007fr,Donato:2003xg}.

The flux of a cosmic ray species $cr$ at the earth location is obtained by convoluting the Green function of the propagation equation
with the source term ${\cal Q}$, given by
\begin{equation}
  \label{eq:source}
  {\cal Q}= BF \frac{\langle \sigma v \rangle \rho^2}{2 m_{DM}^2} \times \sum_i
  \frac{dn^i_{cr}}{dE} BR_i \quad .
\end{equation}
It is worth emphasizing that the astrophysical boost factor in this equation is in general energy dependent. 
The imprint of the clumpiness of the DM halo on the CR spectrum indeed depends on the typical diffusion length, which itself
depends on the injection energy. It has been shown~\cite{Lavalle:2006vb,Lavalle:1900wn} that only high energy positrons and low energy antiprotons
(compared to the injection energy) can be sensitive to a variation of the local DM density. For the rest of the spectrum,
the astrophysical BF never exceeds one order of magnitude.

In this work, the propagation has been carried out by the code DarkSUSY~\cite{gondolo-2004-0407}.
The (default) propagation models implemented in this package correspond to a simplified version of Eq.~(\ref{eq:propag_crs}),
where only the most relevant processes specific to each CR species are included. 
Finally the effect of the solar wind on charged particles is taken into account by applying the force-field approximation,
which results in a shift in energy between the interstellar spectrum (IS) and the one at the top of the atmosphere ($\oplus$),
\be
E_{IS}= E_{\oplus} + |Ze| \phi \quad ,
\ee
and a  depletion of the flux at low energies (below $\sim$ 10 GeV) 
\be
\label{eq:perko}
\frac{d \Phi_{\oplus}}{d E_{\oplus}}= \frac{p^2_{\oplus}}{p^2_{IS}}\frac{d \Phi_{IS}}{d E_{IS}} \quad ,
\ee
where ${p_{\oplus}}$ and ${p_{IS}}$ are the momenta at the Earth and at the heliospheric boundary, $Z e$ is the charge of the CR particle
and  $\phi$ is the solar modulation electric potential  which we took equal to 600 MV (see e.g.~\cite{Bergstrom:1999jc} and reference therein.).

\subsection{Doublet}

The annihilation cross-sections of $H_0$ into $Z Z$, $W^+ W^-$, and $h h$ pairs are given by 
\bea
\sigma(H_0 H_0\rightarrow ZZ)v &=& \frac{g^4}{128\pi c_w^4 m_{H_0}^2}
+\frac{\lambda^2_{A_0}}{16\pi m_{H_0}^2} \nonumber \\
\sigma(H_0 H_0\rightarrow W^+W^-)v &=& \frac{g^4}{64\pi c_w^4m_{H_0}^2}
+\frac{\lambda^2_{H_c}}{8\pi m_{H_0}^2} \nonumber \\
\sigma(H_0 H_0\rightarrow hh)v &=& \frac{\lambda^2_{H_0}}{16\pi m_{H_0}^2} \quad .
\eea
To estimate the various indirect detection signals, we consider seven benchmark points that lead to the relic density required by WMAP
Each of the first six points maximizes the branching ratio in one of the three possible dominant annihilation channels of $H_0$,
for two values of the DM mass $m_{H_0}=1$~TeV and $m_{H_0}=10$~TeV. 
These benchmark points serve to evaluate the spread due to parameters on the particle physics side.
However, the point IV does not satisfy the stability conditions of the potential. 
If these were taken into account (point VII), the maximum branching ratio into $h h$ would be around $25\%$. 
For this last point, the annihilation cross-section is significantly higher. It corresponds to
the degenerate limit situation $m_{H_0} \simeq m_{A_0} \simeq m_{H_c}$, where coannihilations during the freeze-out epoch are the strongest.
Table~\ref{doubletBR} gives the annihilation cross-section and the branching ratios for the benchmark points.
\begin{table}
\small
\bc
\bt{ccccccc}
\hline \hline
Point & $m_{H_0}$~(TeV) & $\sigma v$~(pb) & BR($Z Z$) ($\%$) & BR($W^+ W^-$) ($\%$) & BR($h h$) ($\%$)& BR($t \bar{t}$) ($\%$)\\ 
\hline 
I & 1.0 & 1.47 & 21.0 & 24.8 & 49.7 & 4.5 \\ 
II & 1.0 & 1.56 & 77.7 & 22.3 & 0 & 0  \\ 
III & 1.0 & 1.76 & 17.7 & 82.3 & 0 & 0 \\ 
IV & 10.0 & 1.28 & 0.2 & 0.3 & 99.4 & 0.1  \\ 
V & 10.0 & 1.35 & 99.75 & 0.25 & 0 & 0 \\ 
VI & 10.0 & 1.64 & 0.2 & 99.8 & 0 & 0 \\ 
VII & 10.0 & 3.0 & 25.0 & 50.0 & 25.0 & 0 \\
\hline \hline 
\et 
\ec 
\caption{{\it Annihilation cross-section and branching ratios of the inert doublet DM candidate in the high mass regime. These benchmark points give 
the correct relic abundance as required by WMAP.}} 
\label{doubletBR} 
\end{table} 

The numbers of photons and of neutrinos from the galactic center can be calculated with Eq.~(\ref{fluxGC}).
We will estimate the flux from a solid angle $\Delta \Omega = 10^{-3}$ corresponding to a cone with an aperture of $2^\circ$.
For photons, a typical experiment like FERMI-LAT (former known GLAST) has an angular resolution $\delta \Omega \simeq 10^{-5}$ and energy thresholds $1 \leq E_\gamma \leq 300$~GeV.
For $m_{H_0} = 1$~TeV, the number of photons per annihilation $25 \leq N_\gamma \leq 40$ is slightly higher for annihilations into $h h$.
For $m_{H_0} = 10$~TeV, $30 \leq N_\gamma \leq 100$ is highest for annihilations into $W^+ W^-$.
If we consider a cuspy halo profile like the NFW (Navarro-Frenk-White) one, $\bar{J} \sim 1.3 \cdot 10^3$, which gives
\be 
\Phi_\gamma(\Delta \Omega = 10^{-3}) \simeq {\cal O}(1) \times 2.3 \cdot 10^{-10} 
\left( \frac{m_{DM}}{1~TeV} \right)^{-2} \;\; [{\rm ph \, cm^{-2} \, s^{-1}}] \quad , 
\ee 
which has to be compared to the FERMI-LAT  sensitivity at about $10^{-10} \; {\rm ph \, cm^{-2} \, s^{-1}}$ 
for point sources~\cite{2008cosp...37.2028M,Baltz:2008wd}.
For a flatter profile like the isothermal one ($\bar{J} \simeq 25$), the signal would be broadly distributed over the bulge region.
Even if the total flux lies above the sensitivity for a diffuse flux at about $10^{-10} \; {\rm ph \, cm^{-2} \, s^{-1} \, sr^{-1}}$~\cite{Baltz:2008wd}, 
it would be difficult to disentangle the DM signal from the astrophysical background.

For neutrinos, we consider the experiment Antares and the forthcoming extension KM3net, for which the galactic center is visible.
The energy threshold is $E^\nu \geq 100$~GeV, and the typical angular resolution for $1 \leq E^\nu \leq 10$~TeV is 
$\delta \Omega \simeq 10^{-4}$~\cite{Carr:2007zc}.
The number of neutrinos $N_\nu$ is suppressed in the case of annihilations into $h h$ and their energy spectrum is softer.
For annihilations into $Z Z$ or $W^+ W^-$, $N_\nu \simeq 10$ (1.0) for $m_{H_0}=10$~TeV and $E_\nu^{min} = 100$~GeV (1~TeV).
Using again a NFW profile, we get a flux 
\be 
\Phi_\nu(\Delta \Omega = 10^{-3}) \simeq {\cal O}(1) \times 1.5 \cdot 10^{-12} 
\left( \frac{E_\nu^{min}}{100~GeV} \right)^{-1} \;\; [{\rm \nu \, cm^{-2} \, s^{-1}}] \quad , 
\ee 
while the point source sensitivity of KM3net for 1 year will be 
$\simeq 3 \cdot 10^{-10}\;{\rm \nu \, cm^{-2} \, s^{-1}}$ ($3 \cdot 10^{-11}\;{\rm \nu \, cm^{-2} \, s^{-1}}$) 
for $E_\nu^{min}=100$~GeV (1~TeV)~\cite{Carr:2007zc}.
Therefore, the detection of neutrinos from annihilations of our DM candidate in the galactic center looks very difficult. 



\begin{figure}[t]
\vspace{-0.1cm}
\begin{center}
\begin{tabular}{cc}
\hspace*{-0.5cm}
\includegraphics[width=0.5\textwidth]{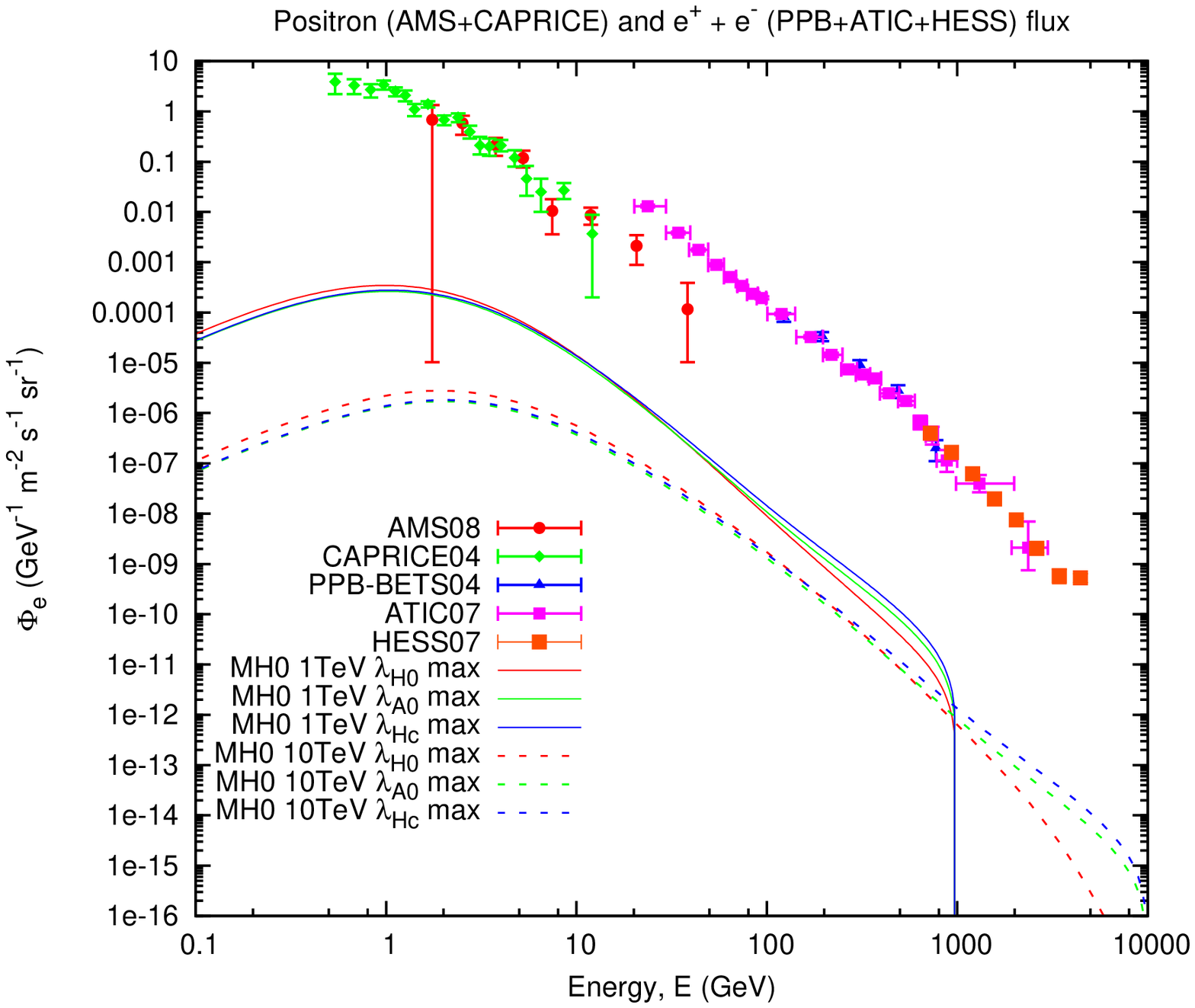} &
\includegraphics[width=0.5\textwidth]{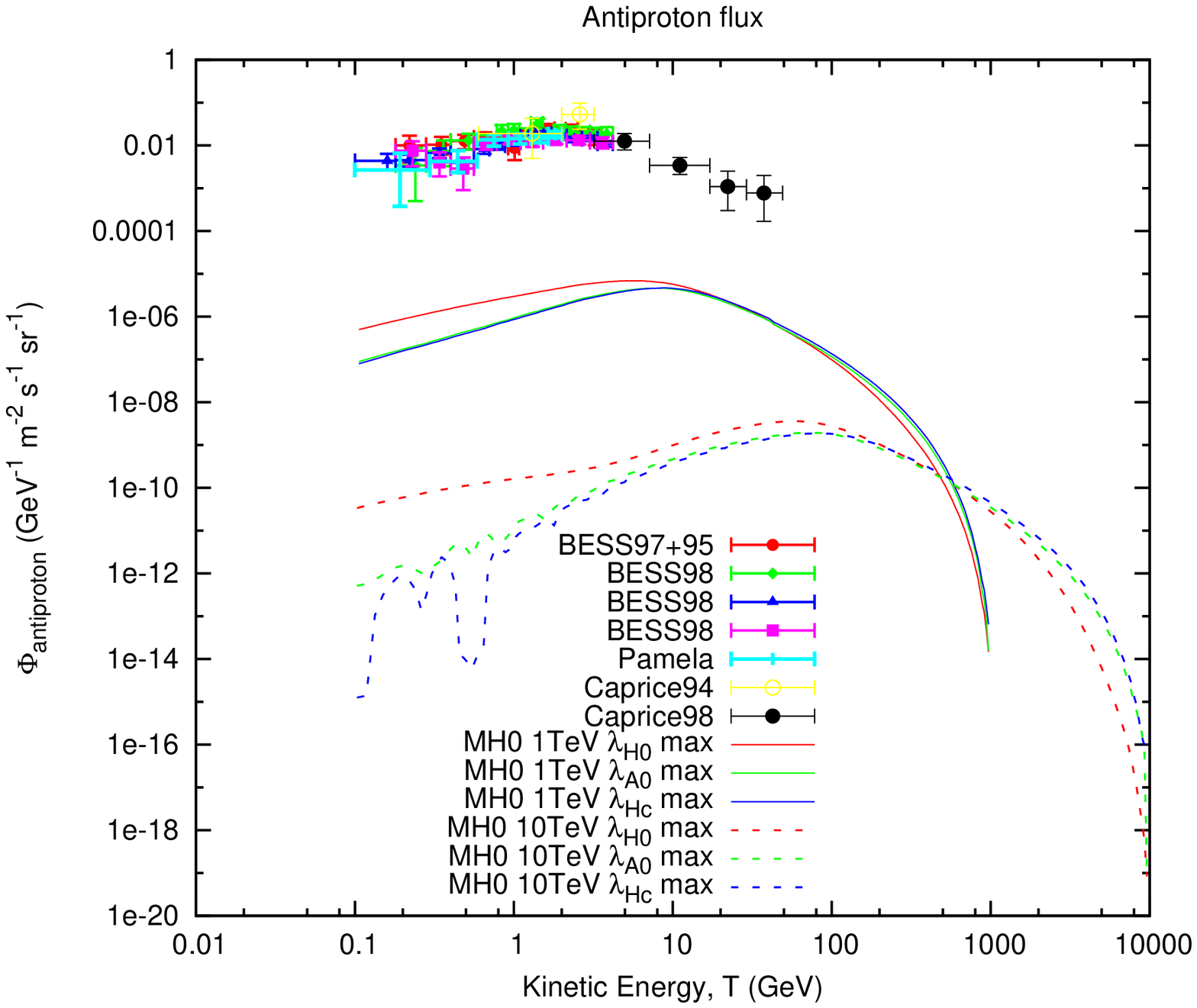} \\ 
\end{tabular} 
\caption{\it  Positron (left panel) and Antiproton (right panel) flux resulting from  $H_0$ annihilation for the
  benchmark models I-VI of Table~\ref{doubletBR}.
  The data are taken from~\cite{Caprice94,Aguilar:2007yf,:2008zzr,Torii:2008xu,Collaboration:2008aaa,Pamelaposflux}
  and from~\cite{Boezio:1997ec,Orito:1999re,Maeno:2000qx,Asaoka:2001fv,Abe:2008sh}.
These figures were obtained with a standard NFW halo profile, a solar modulation potential $\phi= 600$ MV and no boost factor.} 
\label{fig:CRflux} 
\end{center} 
\end{figure}

Positron and antiproton fluxes for the benchmark models of Table~\ref{doubletBR} are shown in Fig.~\ref{fig:CRflux} and we see that they lie well bellow the data.
For $m_{H_0}=1$~TeV, a boost factor of $\sim$ 3 orders of magnitude would be needed to reach the range of the observed positron flux.
In our result, the boosted antiproton flux would however still be below the background.
In the framework of the inert doublet model, such a boost factor is possible only if the DM mass is very close to a resonance. 
The detailed analysis of this possibility is beyond the scope of this paper. 
Going to higher masses decreases the number density of DM particles in the halo. The CR fluxes produced by the annihilation of DM
are therefore even more suppressed for a DM mass larger than 1~TeV.  
For $m_{H_0}=10$~TeV, the positron flux is at least 4 orders of magnitude below the background signal. 
Predictions for the positron fraction in the IDM, as well as a comparison with the excesses seen by PAMELA and ATIC
can be found in Ref.~\cite{Nezri:2009jd}.

Finally, we can notice that the CR spectrums are significantly softer in the case of an annihilation into a pair of Higgs particle.
The Yukawa coupling of $h$ to fermions is directly proportional to the fermion mass, and becomes small compared to gauge couplings for light fermions.
Higgs particles decay mainly into $b \bar{b}$ pairs, leading to multiple hadronization cascades and jets.
The enhancement of the antiproton flux at low energy is particularly clear (1 order of magnitude!).


\subsection{Higher multiplet}

In the case of higher multiplet models, the annihilation cross-sections into $Z Z$, $W^+ W^-$, and $h h$ pairs are given by
\bea
\sigma(\Delta_0 \Delta_0\rightarrow ZZ)v &=& \frac{\lambda^2_3}{64\pi m_0^2} \nonumber \\
\sigma(\Delta_0 \Delta_0\rightarrow W^+W^-)v &=& \frac{(n^2-1)^2g^4}{256\pi m_0^2}
+\frac{\lambda^2_3}{32\pi m_0^2} \nonumber \\
\sigma(\Delta_0 \Delta_0\rightarrow hh)v &=& \frac{\lambda^2_3}{64\pi m_0^2} \quad .
\label{sigmaDI}
\eea
With a large gauge contribution, the $W^+ W^-$ annihilation channel is always dominant.
Omitting the Sommerfeld enhancement, we give in Table~\ref{HMBR} the typical annihilation cross-section
and the branching ratios for a candidate with a mass $m_0=10$~TeV, and the required relic density for $n=3,~5,~7$.
\begin{table}
\small
\bc
\bt{ccccc}
\hline \hline
Model & $\sigma v$~(pb) & BR($Z Z$) ($\%$) & BR($W^+ W^-$) ($\%$) & BR($h h$) ($\%$)\\ 
\hline 
Real Triplet &  2.47 & 24.4 & 51.1 & 24.4 \\ 
Real Quintuplet & 3.81 & 21.7 & 56.5 & 21.7 \\ 
Real Septuplet & 4.19 & 13.1 & 73.7 & 13.1 \\ 
\hline \hline 
\et 
\ec 
\caption{{\it Annihilation cross-section and branching ratios for higher multiplet candidates with a mass $m_0=10$~TeV and a relic density set by WMAP. 
Sommerfeld corrections are omitted.}} 
\label{HMBR} 
\end{table} 
Although the annihilation cross-section increases with the multiplet dimension, the conclusions obtained in the doublet case for the
detectability of the various indirect detection signals still apply.
Gamma ray telescopes offer the most promising search and a better sensitivity than the neutrino detectors.
Again, the production of charged cosmic rays is well below background unless a huge boost factor is applied.

\section{Neutrino masses, leptogenesis and DM at a low scale in the doublet case}\label{sec:NeutrinLepto}

If, to account for the neutrino masses, one adds right-handed neutrinos $N_i$ to the inert doublet model there are 2 possibilities:
either these $N$'s are even under $Z_2$ or they are odd. One interesting consequence of the results obtained above for the doublet case,
which arises for the latter possibility, is that they allow successful generation of both neutrino masses and baryogenesis via leptogenesis
in a way where DM plays an important role. Moreover all three phenomena can be induced
at a scale as low as TeV, even for a hierarchical spectrum of $N$'s.
This has to be compared with the lower bound which exists on the mass of the right-handed neutrinos, $m_N \gtrsim 6\cdot 10^8$~GeV~\cite{Barbieri:1999ma,Hambye:2001eu,Davidson:2002qv},  for a hierarchical spectrum of
$N$'s in the usual type-I seesaw model where only right-handed neutrinos are added to the SM.
To our knowledge, the mechanism we propose in this section is the most simple and minimal way to induce all three phenomena at such a low scale in a related way.

The crucial point at the origin of the fact  that leptogenesis can be generated at such a low scale is that, if the $N$'s are odd under the $Z_2$ symmetry, 
Yukawa coupling involving the $N$'s and the Higgs doublet are forbidden.   The most general lagrangian one can write is \cite{Ma:2006km}
\be
{\cal L} = {\cal L}_{IDM} +i\bar{N}_i \slash {\hspace{-1.8mm} \partial} N_i- \bar{N}_{i} Y_{N_{ij}}  \tilde{H}_2^\dagger L_j -\frac{1}{2} m_{N_i} N_{i} N_{i}
\label{inertseesawL}
\ee
with $\tilde{H}_2=i \tau_2 H_2^{*}$, i.e.~only the Yukawa couplings with the inert doublet are allowed.
As a result neutrino masses cannot be generated at tree-level in the usual way but only at one loop through two DM inert doublets, Fig.~\ref{fig:numassdiagram},
which for $m_{Ni}>> m_{H^0,A^0,H^c}$ gives (at lowest order in $\lambda_5 v_0^2/m_{H_0}^2$)
\be
(m_\nu)_{ij}=-\frac{\lambda_5 v_0^2}{16 \pi^2} \sum_k \frac{Y_{Nki} Y_{Nkj}}{m_{N_k}}\Big[\log \frac{m^2_{H_0}}{ m^2_{N_k}}+1\Big].
\label{mnuinert}
\ee
With respect to the standard tree-level seesaw model, which gives $(m_{\nu})_{ij}= -\frac{v_0^2}{2} \frac{Y_{Nki} Y_{Nkj}}{m_{N_k}}$,
this "radiative seesaw" mechanism leads consequently to an extra suppression of the neutrino mass by a factor
$\frac{\lambda_5 v_0^2}{8 \pi^2} [\log (m^2_{H_0}/ m^2_{N_k})+1]$ for each $N_j$ contribution.

\bfig
\bc
\includegraphics[width=0.28\textwidth]{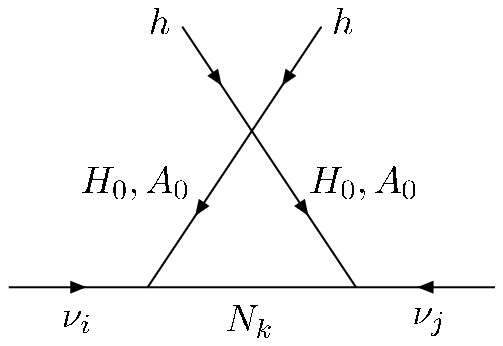}
\caption{One-loop neutrino mass diagram.}
\label{fig:numassdiagram}
\ec
\efig

As for leptogenesis in this framework, it proceeds from the $N \rightarrow L H_2, \bar{L} H_2^*$, that is to say in the same way as in the usual type-I seesaw model,
replacing all ordinary Yukawa couplings to a Higgs doublet by the inert doublet Yukawa couplings of  Eq.~(\ref{inertseesawL}), Fig.~\ref{fig:leptoH2diagrams}.
For the lightest right-handed neutrino $N_1$ and $m_{N_{2,3}} >> m_{N_1}$ this gives the CP-asymmetry
\be
\varepsilon_{N_1}=-\sum_{j=2,3} \frac{3}{16 \pi} \frac{m_{N_1}}{m_{N_j}} \frac{\sum_i Im[(Y_{N{1i}} Y^\dagger_{Nij})^2]}{\sum_i |Y_{N1i}|^2}
\label{epsilonNinert}
\ee
It is the extra suppression above of the neutrino masses versus absence of any extra suppression of the CP-asymmetry which allows to lower the scale of leptogenesis, as we will now show in details by deriving the various relevant bounds:

\paragraph{1.~Leptogenesis and neutrino mass bounds on $\varepsilon_{N_1}$, $m_{N_1}$ and $\lambda_5$.}  In full generality in the type-I seesaw model, $\varepsilon_{N_1}$ is bounded by the size of neutrino masses~\cite{Davidson:2002qv}
\be
|\varepsilon_{N_1}|=\frac{3}{8 \pi} \frac{m_{N_1}}{v_0^2} \frac{|Im[Tr m_\nu^{{1}\dagger} m_\nu^{{2,3}} ]|}{\tilde{m}_1}
\leq \frac{3}{8 \pi} \frac{m_{N_1}}{v_0^2} (m_{\nu_3}-m_{\nu_1})
\label{boundepsilon}
\ee
with $m_\nu^i$ the contribution of $N_i$ to the neutrino mass matrix, $m_{\nu_3}$ ($m_{\nu_1}$) the mass of the heaviest (lightest) neutrino,
and $\tilde{m}_1=\sum_i |Y_{N1i}|^2 v_0^2/2m_{N_1}$.  This leads to the lower bound $m_{N_1} \gsim 6 \cdot 10^8$~GeV, imposing that the baryon asymmetry produced
$n_B/s=-\frac{28}{79} n_L/s= -\frac{135 \zeta(3)}{4 \pi^4 g_\ast} \frac{28}{79} \varepsilon_{N_1} \eta=-1.35 \cdot 10^{-3} \varepsilon_{N_1} \eta$
is at least equal to the WMAP value, $n_B/s=9 \cdot 10^{-11}$ (assuming a maximal efficiency $\eta=1$), with $g_\ast=108.75$
the number of active degrees of freedom at the time of the creation of the asymmetry.
For the radiative seesaw case, taking for simplicity all $[\log (m^2_{H_0}/
m^2_{N_k})+1]$ factors equal to unity, one gets an extra $8\pi^2/\lambda_5$ enhancement factor of the CP-asymmetry and hence a decrease of the $m_{N_1}$ lower bound by the same amount:\footnote{Notice that if one doesn't take the approximation that all logarithmic factors are unity one can get even a much more relaxed bound,
taking  for instance $[\log (m^2_{H_0}/ m^2_{N_k})+1]\simeq 0$. We will not consider this peculiar possibility preferring here to stay generic.}
\be
|\varepsilon_{N_1}|=\frac{8 \pi^2}{\lambda_5}\frac{3}{8 \pi} \frac{m_{N_1}}{v^2} \frac{|Im[Tr m_\nu^{{1}\dagger} m_\nu^{{2,3}} ]|}{\tilde{m}_1}
\lesssim \frac{8 \pi^2}{\lambda_5} \frac{3}{8 \pi} \frac{m_{N_1}}{v^2} (m_{\nu_3}-m_{\nu_1})
\label{boundepsilon}
\ee
\be
m_{N_1}\gtrsim \frac{\lambda_5}{8 \pi^2}\, 6\cdot 10^8\,\hbox{GeV}
\label{mNbound}
\ee
The change of number of effective degrees of freedom due to the extra active inert component(s) is of
little importance and can be neglected.

Consequently for successful leptogenesis the lower bound can be lowered down to any value $m_{N_1}$ if
\begin{equation}
\lambda_5 \lesssim 1.5 \cdot 10^{-4} \cdot (m_{N_1}/1\,\hbox{TeV})\,\,\,\,\, \leftrightarrow \,\,\,\,\,\, m_{A_0}-m_{H_0}< 9\,\hbox{MeV} \cdot (500\,\hbox{GeV}/m_{\rm DM})\cdot (m_{N_1}/1\, \hbox{TeV})
\label{mN1bound}
\end{equation}

\paragraph{2.~Bound on $\lambda_5$ from DM constraints in the low mass regime and corresponding lower bound on $m_{N_1}$ for successful leptogenesis.}

In the low mass DM regime, where
$m_{\rm DM}< m_W$,\footnote{Leptogenesis in this case has been considered in Ref.~{\cite{Ma:2006uv}} for large right-handed neutrino masses around $10^9$~GeV.} to avoid too fast $H_0$-$A_0$ coannihilation leading to too low relic density, it is necessary that
the mass splitting is large enough, $m_{H_0}-m_{A_0}>7$~GeV \cite{Barbieri:2006dq}, or equivalently
$\lambda_5\gtrsim 1.6\cdot 10^{-2}$.
This bound is incompatible with successful leptogenesis, i.e.~Eq.~(\ref{mN1bound}), unless
\begin{equation}
m_{N_1}^{Low\,regime}\gtrsim  110~\hbox{TeV}
\end{equation}
which is interestingly low but not enough to be reachable in a not too long term at colliders.

\bfig
\bc
\includegraphics[width=0.45\textwidth]{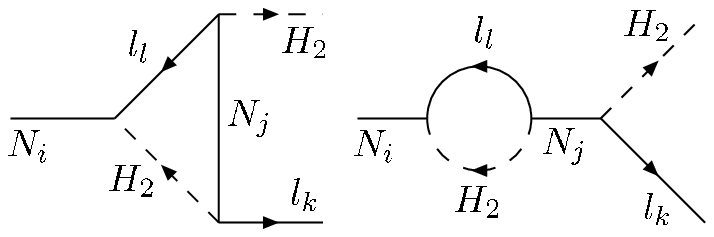}
\caption{One-loop diagrams contributing to the asymmetry
from the $N_i$ decay, involving the DM inert doublet.}
\label{fig:leptoH2diagrams}
\ec
\efig

%

\paragraph{3.~Lower bound on $m_{N_1}$ for successful leptogenesis in the high mass regime.}
In the high mass regime, as we have seen in Section.~\ref{subsec:quartcoupl}) and unlike in the low mass regime, the relic density constraint doesn't lead to any lower or upper bound on $\lambda_5$ (apart from a perturbativity bound).
As explained above the only lower bound comes from direct detection constraint, $m_{A_0}-m_{H_0} \gtrsim 100$~keV, which is well below the bound of Eq.~(\ref{mN1bound}) (unless $m_{N_1}\lesssim 20$ GeV, that is to say well below the sphaleron decoupling temperature anyway).
The only relevant constraint in this case comes therefore from successful conversion of the lepton asymmetry into a baryon asymmetry by the sphalerons. Assuming a sphaleron decoupling scale of order $150$ GeV\cite{Burnier:2005hp} and taking into account that the creation of the lepton asymmetry cannot be instantaneous, one gets the constraint:
\be
m_{N_1}^{High\,regime} \gtrsim 1\,\hbox{TeV}
\ee

\paragraph{4.~Bounds on the $N_i$ Yukawa couplings.}
All bounds above are obtained assuming that there is no efficiency suppressions of the lepton asymmetry produced. This requires that the decay width of $N_1$ satisfies the out of equilibrium condition:
\be
\Gamma_{N_1}=\frac{1}{8 \pi} |Y_{N_{1j}}|^2 m_{N_1} < H(T=m_{N_1})\simeq \sqrt{\frac{4 \pi^3 g_\ast}{45}}\, \frac{T^2}{M_{ Planck}}\Big|_{T=m_{N_1}}
\ee
which gives
\be
|Y_{N_{1j}}|^2< 4 \cdot 10^{-14} \cdot (m_{N_1}/1\,\hbox{TeV})
\label{YN1bound}
\ee
Notice that the smallness of these couplings
doesn't induce any suppression of the CP-asymmetry because these couplings essentially cancel in it (up to phases), see Eq.(\ref{epsilonNinert}). Eq.~(\ref{YN1bound}) implies that $N_1$ gives neutrino mass contributions smaller than the atmospheric or solar mass splittings. These splittings must therefore be dominated by the contribution of $N_{2,3}$ and the neutrino mass spectrum is necessarily hierarchical. The heavier $N_{2,3}$ states must have necessarily larger Yukawa couplings.
Eq.~(\ref{mN1bound}), together with the lower bound $m_{\nu_3}\geq \sqrt{\Delta m^2_{atm}}=0.06$~eV
and Eq.~(\ref{mnuinert}) imply a lower bound on the $N_{2}$ or $N_3$ Yukawa coupling
\be
Y_{Nji}\, \gsim \,1 \cdot 10^{-3}  \cdot \Big(\frac{m_{N_j}}{m_{N_1}}\Big)^{1/2} \quad \quad (j=2\,\hbox{or}\,3)
\label{YN23bound}
\ee
for at least one lepton flavor $i$ value. This bound can also be obtained directly from Eq.~(\ref{epsilonNinert}) imposing that the CP-asymmetry is large enough to induce the observed baryon asymmetry.

Comparing Eq.~(\ref{YN1bound}) with Eq.~(\ref{YN23bound}), one therefore concludes that to induce leptogenesis at a low scale in a non-resonant
way \cite{Frigerio:2006gx,Boubekeur:2004ez,Hambye:2001eu} one needs a hierarchical structure of Yukawa coupling
(similar to the one of the charged leptons).

Eq.~(\ref{YN23bound}) gives Yukawa coupling values much larger than in the
usual type-I seesaw model where in order to give neutrino masses $\sim \sqrt{\Delta m^2_{atm}}$ one needs
Yukawa couplings 
$\sim {\cal O }(10^{-6})$ for $m_N\sim 1$~TeV (unless cancellations occur between the Yukawa couplings in the neutrino masses).

\paragraph{5.~$\Delta L =2$ washout constraint.} Finally in order to have an efficiency of order unity as assumed above it is also necessary that there is no washout from scattering processes. The most dangerous are the $N_{2,3}$ mediated $\Delta L=2$ ones, due to the fact that the $N_{2,3}$  Yukawa couplings must be fairly large.
However for values of $Y_{N_{2,3}}$ of the order of the bounds in Eq.~(\ref{YN23bound}), it can be checked from the Boltzmann
equation~\cite{Frigerio:2006gx,Boubekeur:2004ez,Hambye:2001eu}
that this effect is moderate or even negligible (even without needing to play with flavor effects which could be invoked otherwise to suppress further this $\Delta L =2$ washout effect).\\

Taking into account all constraints above, as a numerical example\footnote{
  for simplicity we assume in the numerical example that $N_3$ is heavier and has little effect on leptogenesis},
for $\lambda_5=10^{-4}$, $m_{N_1}=2$~TeV, $m_{N_2}=6$~TeV, $Y_{Ni}\simeq
\hbox{few} \,10^{-8}$, max$_i(Y_{N2i})\simeq 4\cdot 10^{-3}$ and $m_{H_0}$
above~$\sim~510$~GeV and sizably below $m_{N_1}$ we get $m_{\nu_3}=\sqrt{\Delta m^2_{atm}}$, 
the WMAP value $n_b/s\simeq 9 \cdot 10^{-11}$ (with no sizeable suppression of the efficiency)
and a dark matter relic abundance which can be easily consistent with the WMAP range above (i.e.~in agreement with the results of
Section.~\ref{subsec:quartcoupl}).
An interesting property of this framework is that it involves Yukawa coupling of $N_2$ and/or $N_3$ much larger than in the usual seesaw model.
Unlike in the latter case, it is therefore conceivable in the not too long
term to produce the  right-handed neutrinos of the present extended IDM at colliders.
 This is an interesting phenomenological possibility to test in addition to the nature of DM, the related origin of neutrino masses and baryogenesis via leptogenesis.\footnote{Note finally that in this framework the $N's$ produce not only a $L$ asymmetry but also a inert doublet asymmetry, but for $\lambda_5$ satisfying the direct detection lower bound above this asymmetry rapidly
is washed away by the $\lambda_5$ driven $H_2 H_2 \leftrightarrow H_1 H_1$ processes together with pure Higgs boson self interactions.}

\section{Summary}
\label{sec:finale}

Properties of scalar DM candidates with $SU(2)_L$ quantum numbers are  
driven by their known gauge interactions and by their scalar quartic  
interactions. If the quartic couplings are not much smaller than the  
gauge interactions their effects cannot be neglected. This leads, for  
each model, to a range of values of DM masses which can reproduce the  
observed DM relic density. Allowing for 3$\sigma$ uncertainty on the WMAP DM abundance,  the lower  edges of these ranges are given  
to a good approximation by the value obtained without scalar  
interactions: $0.51$~TeV, $1.7$~TeV, 4.4~TeV, 7.6~TeV ($0.51 
$~TeV, $2.2$~TeV, 9.0~TeV, 21.4~TeV), without (with) Sommerfeld  
corrections for $n=2$ complex and $n=3,5,7$ real multiplets respectively. The  
upper bound lies from 16 TeV to 60 TeV depending on the model and the  
perturbativity condition one assumes. For a complex multiplet with  
$n=3,5,7$ all these values have to be reduced by a $\sqrt{2}$ factor.

These models are quite predictive.
For the inert doublet model, since there are three relevant quartic  
couplings, for a fixed value of the DM mass, there is a two  
dimensional space of values of the quartic couplings that reproduces  
the observed relic density (see section~\ref{subsec:quartcoupl}). Its  
shape is close to the one of an ellipsoid.
This leads to an upper bound on each quartic coupling and therefore  
on the inert Higgs doublet component mass splittings, given in Fig.~ 
\ref{lambdamax}. As these upper bounds have an asymptotic behaviour  
for large DM masses, there exists an absolute maximum upper bound on  
each one, from 13~GeV to 17~GeV depending on the splitting considered.

For the multiplets of dimension $n \gtrsim 3$, as shown in section~ 
\ref{sec:highermultipletmodel}, there is only one relevant quartic  
coupling (the $H_n^\dagger H_n H_1^\dagger H_1$ coupling).
Therefore the value of this quartic coupling is fixed by the mass of  
the DM, see Fig.~\ref{fig:M0delambda3}. No mass splitting between the  
multiplet components greater than $[(n-1)/2]\times 166~\mathrm{MeV}$  
are generated in this case.

For the doublet case the direct detection elastic cross sections can  
be enhanced by up to 2 orders of magnitude by the scalar coupling  
contribution, see Fig.~\ref{dddmax}. The cross section is predicted  
to lie within the range $[0.1\--9.0]\cdot 10^{-9}$~pb. Consequently  
they can exceed the sensitivity of future planned experiments by a  
similar amount. For higher multiplets the cross section is completely  
fixed by the DM mass, leading to values which can be ruled out by  
future experiments such as Xenon1T, from $0.9 \cdot 10^{-9}$~pb to  
$40 \cdot 10^{-9}$~pb.
Indirect detection signals can also be enhanced by the quartic  
coupling contributions. For a standard Navarro-Frenk-White halo  
profile, and without any boost factor, the total gamma ray flux is  
within reach of future telescope such as FERMI-LAT.
Search for high energy neutrinos from the galactic center is  
complementary to the gamma ray signal. However, without boost, the  
predicted flux is 1-2 orders of magnitude below the projected  
sensitivity of $km^3$ size detectors.
The antiproton and positron fluxes are 3-4 orders of magnitudes below  
the expected background, see Fig.~\ref{fig:CRflux}. Very large boost  
factors would be therefore necessary to have a signal exceeding this  
background. Since the scalar DM models are viable over large ranges  
of masses, it is clear that some values of the mass within these  
ranges will lie on the top of a Sommerfeld resonance, possibly leading to a  
large boost for all indirect detection signals. A precise  
determination of these resonances potentially relevant for PAMELA and ATIC
experiment results is beyond the scope of this paper.

The values of DM masses we have obtained for the higher multiplet are  
clearly too high to allow DM particle production at the LHC collider.  
It would however be very interesting to analyze the  
possibilities to produce the inert doublet components (in particular  
the charged ones) in the range $\sim 0.5-1.5$ TeV.

Finally if one adds right-handed neutrinos to the doublet model it is  
possible to successfully generate in a simple way the neutrino  
masses, the baryon asymmetry of the universe (via leptogenesis in a  
non-resonant
way) and the dark matter relic density, all this in a related way and  
at a scale as low as TeV.
Conversely this means that if a second Higgs doublet is added to the ordinary type-I  
seesaw model
one can lower the right-handed neutrino masses and therefore the  
leptogenesis scale from $10^9$~GeV down to TeV, and at the same time  
explain the observed DM relic density with a TeV scale scalar  
candidate from the second Higgs doublet.

\section*{Acknowledgments}

The authors would like to thank C. Arina, J.-M. Fr\`ere and M. Tytgat for helpful discussions.
We thank Alexander Pukhov and Genevi\`eve B\'elanger for clarifications about MicrOMEGAs.
We are also indebted to C. Yaguna for his help in handling DarkSUSY.
This work is supported by the FNRS-FRS and the IAP belgian funds. L.L.H. receives financial support through a postdoctoral fellowship of
the PAU Consolider Ingenio 2010 and was partially supported by CICYT
through the project FPA2006-05423 and by CAM through the project
HEPHACOS, P-ESP-00346.

\newpage

\appendix

\section{Higher $SU(2)_L$ multiplets}

\subsection{Generators, real and complex multiplets}\label{sec:annexGener}

In this section, we give explicit expressions for the $SU(2)$ generators in the representation $\mathbf{n}$,
and define real and complex multiplets. 

Let us consider a multiplet $H_n$ in the representation $\mathbf{n}$ 
of $SU(2)_L$. The generators must be chosen and normalized
so as to satisfy the commutation relations of $SU(2)$,
$[\tau_a^{(n)},\tau_b^{(n)}]=i \epsilon_{abc}\tau_c^{(n)}$, but
their form is not uniquely determined.
The \emph{spherical basis}, where the 
third generator is diagonal
\be
\label{eq:generatorspheric3}
\tau^{(n)}_3=\mathrm{diag}(j_n,j_n-1,\dots,-(j_n-1),-j_n)) \quad ,
\ee
with $j_n=(n-1)/2$, is particularly convenient as the 
components of the multiplets are eigenstates of the 
electromagnetic charge.
The first two generators can be constructed from the ladder
operators $\tau_\pm^{(n)} =\tau^{(n)}_1\pm i \tau^{(n)}_2$
which are given by
\be
\label{eq:generatorsphericpm}
\tau_+^{(n)}|e_k^{(n)}\rangle = \left\{
\begin{array}{lcl}
-[(j_n-k)(j_n+k+1)]^{1/2}|e_{k+1}^{(n)}\rangle & , & k \geq 0 \\\nonumber
\\\nonumber
[(j_n-k)(j_n+k+1)]^{1/2} |e_{k+1}^{(n)}\rangle & , & k < 0
\end{array}
\right.
\ee
and $\tau^{(n)}_-= \left(\tau^{(n)}_+\right)^T$, where 
$|e_k^{(n)}\rangle$ (with $k=-j_n,\,-j_n+1,...,\,j_n$)
are the basis-vectors. As an example, for the triplet case,
\be
\label{eq:generatortripletspheric}
\tau_1^{(3)}=\frac{1}{\sqrt{2}}\begin{pmatrix} 0 & -1 & 0 \\ -1 & 0 & 1 \\
0 & 1 & 0
\end{pmatrix} \quad
\tau_2^{(3)}=\frac{1}{\sqrt{2}}\begin{pmatrix} 0 & i & 0 \\ -i & 0 & -i \\
0 & i & 0  \end{pmatrix} \quad
\tau_3^{(3)}=\begin{pmatrix} 1 & 0 & 0 \\ 0 & 0 & 0 \\ 0 & 0 & -1 \end{pmatrix}~.\\
\ee
These generators are related to the generators $({\tau'_a}^{(3)})^{bc}=
-i\epsilon^{abc}$ in the \emph{cartesian basis} by the unitary 
transformation matrix
\be
\label{eq:transfobasistriplet}
U=\frac{1}{\sqrt{2}}\begin{pmatrix} 1 & 0 & 1
\\ i & 0 & -i \\ 0 & \sqrt{2} & 0
\end{pmatrix}~.
\ee

From a group theory point of view, any representation of $SU(2)$ is real,
in the sense that it is equivalent to its complex conjugate. For the
representation $\mathbf{n}$ of $SU(2)$, the matrix $T_n$ which realizes this
equivalence,
\be
T_n \tau_a^{(n)} T_n^{-1}=-\tau_a^{(n)*}
\ee
is given in the spherical basis by
\be
T_n |e_k^{(n)}\rangle = \left\{
\begin{array}{lcl}
(-1)^{n+1}|e_{-k}^{(n)}\rangle & , & k \geq 0 \\\nonumber
\\\nonumber
|e_{-k}^{(n)}\rangle & , & k < 0
\end{array}
\right.
\ee
Therefore, for a multiplet $H_n$, the conjugate multiplet
$\tilde{H}_n \equiv T_n H_n^*$ also transforms as the
representation $\mathbf{n}$ under $SU(2)$. 

We can distinguish between real multiplets for which
$\tilde{H}_n=H_n$ and complex ones for which $\tilde{H}_n\neq H_n$. 
A complex multiplet of dimension $n$ contains twice as many degrees of
freedom as a real multiplet of same dimension.
In the spherical basis given above, the components of charge $Q$ and $-Q$ of real multiplets
are complex conjugate of one another, in the cartesian basis, all components
are real fields. Finally, it is to check that $H_n^\dagger\tau_a H_n$ vanishes
for real multiplets.


\subsection{Complex multiplets: mass spectrum and comparison to the real case}\label{sec:complexmultiplet}
In the section \ref{sec:highermultipletmodel}, we presented in
details the mass-spectrum and properties of real multiplets. 
In this section we analyze whether a complex multiplet can
still be considered as a minimal candidate of dark matter, and how
its phenomenology differs from the real case. 
In the spherical
basis where the third generator is diagonal, the complex multiplet
with $Y=0$ can be cast in the following form
\be
H_n =\begin{pmatrix} \Delta_1^{(j_n)} \\ \dots \\
\Delta^{(0)}\\ \dots \\
\Delta_2^{(-j_n)}
\end{pmatrix} \quad ,
\label{CMdec}
\ee
where the upper index of a component corresponds to its electric charge.
Notice that the number of independent fields has doubled compared to the real case
(which explains the absence of a normalization factor $1/\sqrt{2}$ in Eq.~(\ref{CMdec})).

In the case of a complex $H_n$ with $Y=0$, the lagrangian and
potentials given in
Eqs.~(\ref{eq:lagrangianmultiplet}-\ref{eq:potentialmultiplet})
are not the most general ones anymore, because of the possibility
of mixed products between $H_n$ and $\tilde{H}_n$ (like
$(H_n^\dagger \tilde{H}_n)$ or $(H_n^\dagger \tau_a^{(n)}
\tilde{H}_n)$). However, the complex multiplet can be decomposed
into two real multiplets. Indeed, if we define 
\bea
A_n &=& \frac{1}{\sqrt{2}}(H_n + \tilde{H}_n) \nonumber \\
B_n &=& \frac{i}{\sqrt{2}}(\tilde{H}_n - H_n) \quad ,
\label{realdec}
\eea
it is easy to check that $A_n$ and $B_n$ are real multiplets, that
is $A_n = \tilde{A}_n$ and $B_n = \tilde{B}_n$. Therefore the most
general model with a complex multiplet $H_n$ with vanishing
hypercharge is equivalent to a model with two interacting real
multiplets $A_n$ and $B_n$. Following the minimality criterium
that enables us to make a systematic study, we will not pursue the
details of such a model.

There is however one case where the decomposition of
Eq.~(\ref{realdec}) is forbidden, namely when $H_n$ is charged
under some additional gauge group $U(1)_{Q'}$. In that case, the
potential of Eqs.~(\ref{eq:potentialmultiplet}) is still the most
general one and the  $\lambda_4$ and $\lambda_5$ terms do not
vanish. In particular, $\lambda_5$ generates a mass splitting at
tree-level between the components of $H_n$,
\be
m^2 ({\Delta_\alpha^{(Q)}}) = \mu^2 + \frac{\lambda_3 v_0^2}{2} + (-1)^\alpha Q
\frac{\lambda_5 v_0^2}{4} \quad , 
\label{newmrel}
\ee
%
with $Q=1, \dots ,\,j_n$ and $\alpha=1,2$. As a consequence, half of the charged
fields of the multiplet are lighter than the neutral component at
tree-level, the latter cannot therefore be a DM candidate, which
rules out the model. At one-loop however, one has to take into
account the additional splitting generated by the coupling to
gauge bosons
\be
m\left({\Delta_\alpha^{(Q)}}\right)-m\left({\Delta^{(0)}}\right) =
Q^2 \Delta M_g \quad , 
\ee 
with $\Delta M_g \simeq (166 \pm 1)~\mathrm{MeV}$. 
Therefore, the neutral component stays the lightest particle as long as
\be \lambda_5 \; \lesssim \; 2.2\times
10^{-2}\left(\frac{m_0}{1~\mathrm{TeV}}\right)~,
\label{smalll5} \ee
where $m_0\equiv m\left({\Delta^{(0)}}\right)$.
In this calculation, scalar sector induced 1-loop corrections 
have been neglected for the following reason. Only $\lambda_5$ 
induces charge-dependent 1-loop corrections to the scalar mass 
(via loops of the SM Higgs)
in the scalar sector so all corrections are proportional to 
$\lambda_5$, $\lambda_5^2$ or $\lambda_3 \lambda_5$. In the first 
two cases, it is clearly impossible for 1-loop corrections to 
compensate tree-level splittings because they are proportional to the 
same coupling. In the last case also, $\lambda_3$ would have to be 
taken far beyond the perturbative regime for 1-loop corrections 
to be non-negligible.

The constraint Eq.~(\ref{smalll5}) puts a strong upper bound on
the only coupling that could impact the DM phenomenology of the
complex models compared to the real cases since $\lambda_4$
doesn't introduce new couplings to the SM particles. As a result,
the complex cases are mostly equivalent to the real cases, except
for the fact that the number of degrees of freedom has doubled. As
a consequence, the total cross-section of (co)annihilation increases by
a factor 2. As $\Omega_{\rm DM}\propto m_0^{-2}$ (see 
section~\ref{sec:relic}), for a given relic
density, the mass of the DM candidate is smaller by a factor
$\sqrt{2}$ in the case of a complex multiplet compared to the
corresponding real case.

\section{Feynman Diagrams for Annihilation and Coannihilation in Inert Multiplet Models}\label{sec:app:diagIDM}

\subsection{Inert Doublet Model}
We represent below the annihilation and coannihilation
processes. They are organized by types of output particles.

\bfig[ht]
\bc
\includegraphics[width=0.95\textwidth]{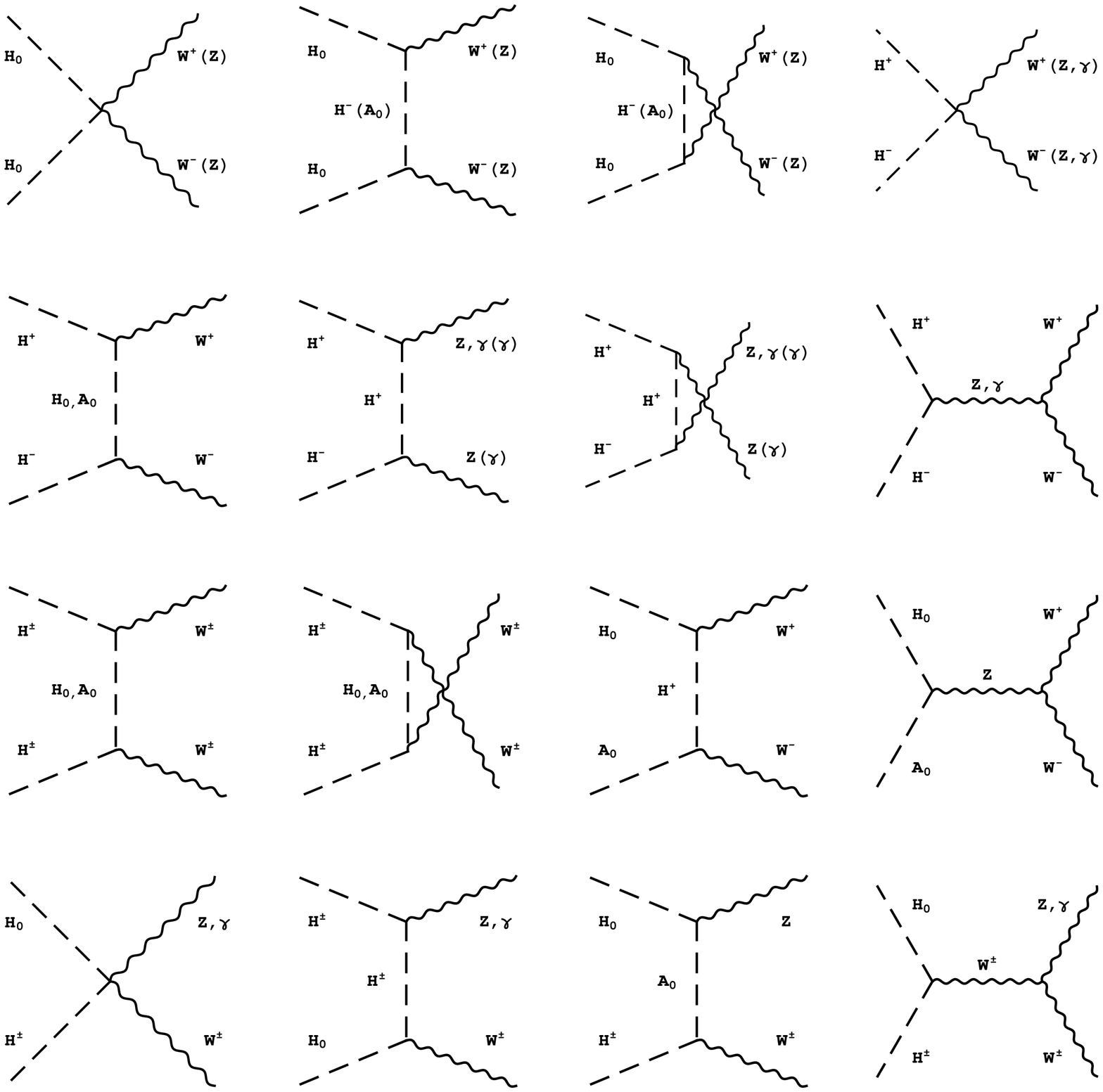}
\caption{Pure gauge (co)annihilation channels}
\label{GDfig}
\ec
\efig
\newpage
\bfig[hbt]
\bc
\includegraphics[width=0.95\textwidth]{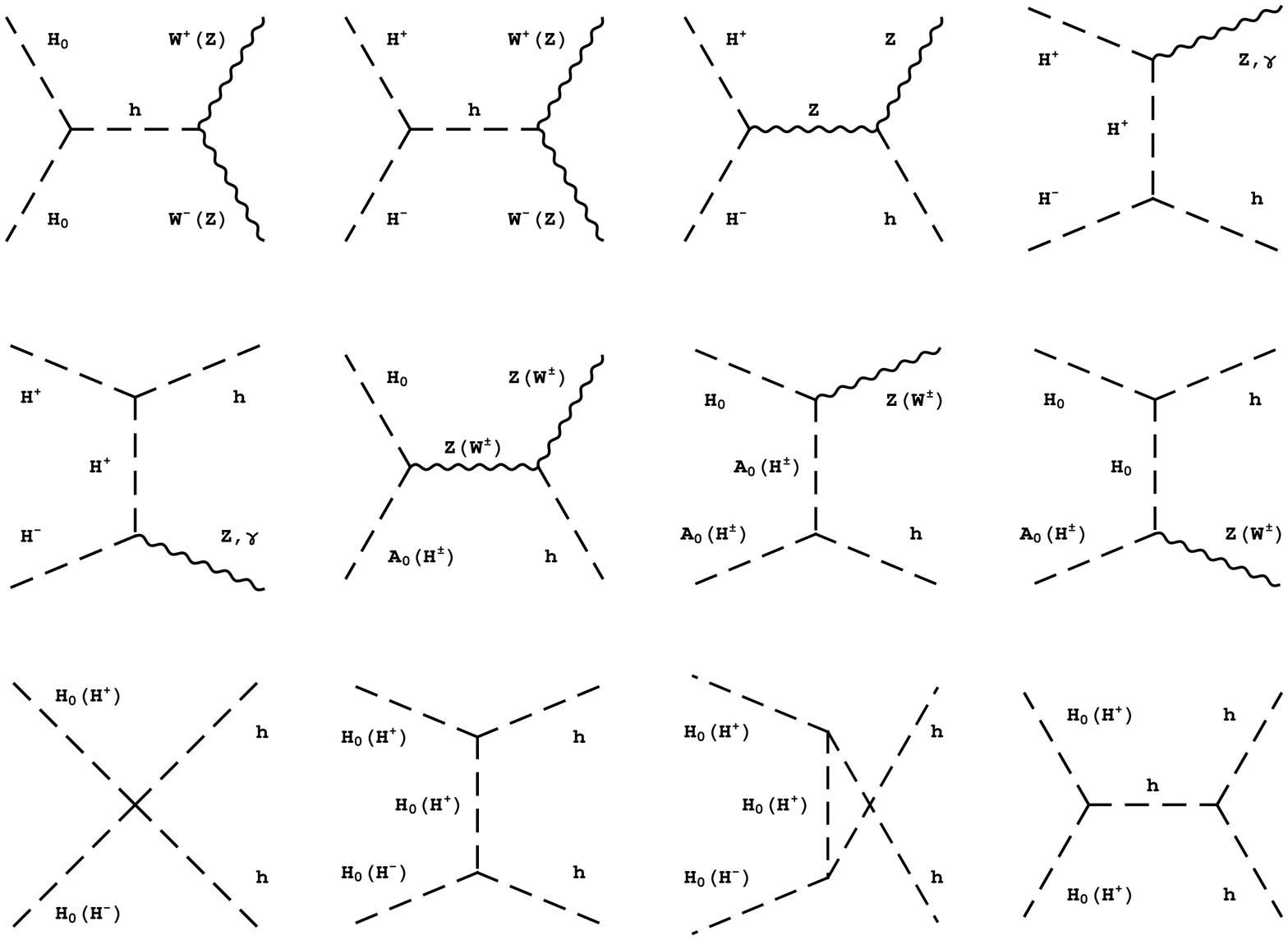}\\
\caption{(Co)Annihilation channels involving the usual Higgs particle for the doublet model.}
\label{HDfig}
\includegraphics[width=0.95\textwidth]{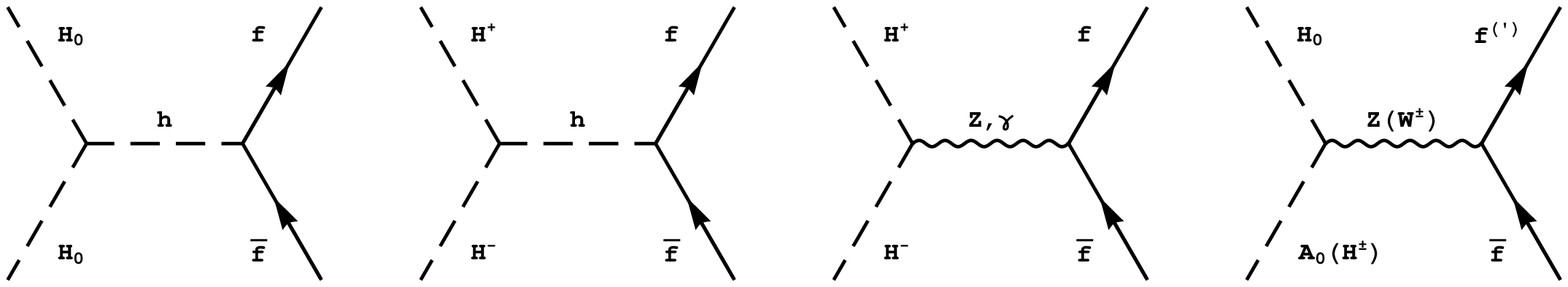}
\caption{(Co)Annihilation channels into fermions for the doublet model.}
\label{FDfig}
\ec
\efig


\subsection{Inert Multiplet Model}\label{sec:Feynmangraphmultiplet}
Like for the doublet, we represent below the annihilation and
coannihilation processes for the inert multiplet model of
any dimension $n$. The contributions are organized by type of
output particles: 
Figs.~\ref{fig:Feynmangraphmultiplet2},\ref{fig:Feynmangraphmultiplet3},  and \ref{fig:Feynmangraphmultiplet1}
represent the channels involving Higgs particles, the
fermionic channels and the pure gauge channels respectively. 

One diagram encloses several cases corresponding to all the
possible values of  $Q$ which stands for the absolute value of the charge of the $\Delta^{\pm
  Q}$ $n$-uplet component. Remember that $Q=0,1,\dots,j_n$, with $j_n=(n-1)/2$.
Notice that some diagrams do not exist for all
charges or all models. For example, those involving $\Delta^{Q-1}$ cannot be applied to the $Q=0$
case and those involving $\Delta^{Q+2}$ do not exist in the triplet case.
Moreover, some interactions are not possible because of the absence of coupling. 
This is the case  of {\it e.g.} the sixth diagram  of
Fig.~\ref{fig:Feynmangraphmultiplet1} for $Q=0$ because $\Delta^0$ does not
couple to $Z$ or the photon.
%

\bfig
\bc
\includegraphics[width=0.9\textwidth]{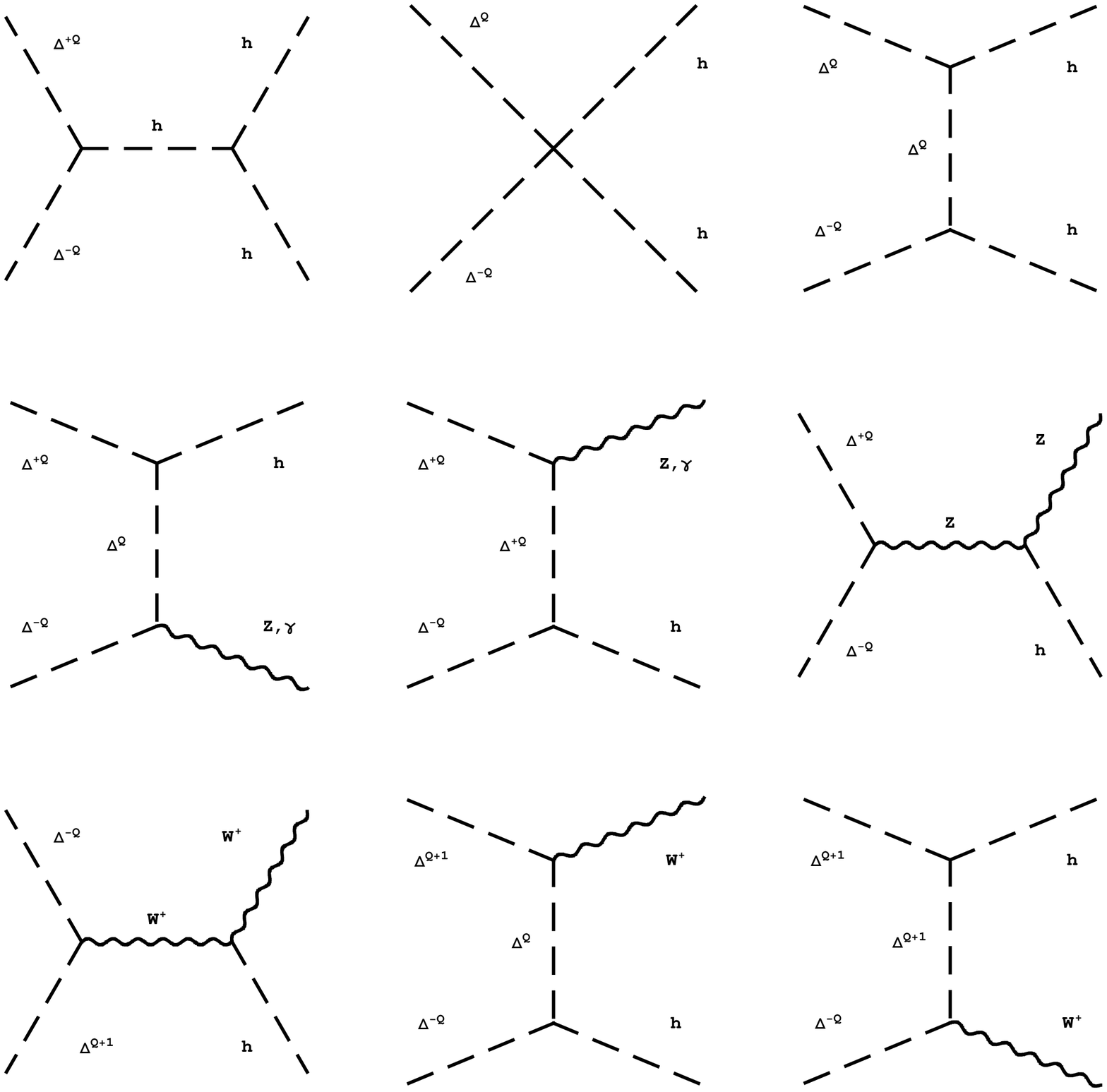}
\caption{(Co)Annihilation channels involving the SM Higgs for the
Multiplet model.}
\label{fig:Feynmangraphmultiplet2}
\ec
\efig

\bfig
\hspace*{-1.5cm}\includegraphics[width=1.3\textwidth]{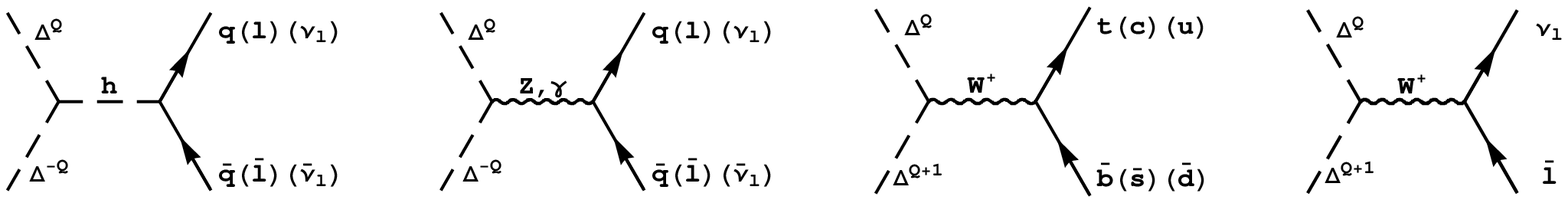}
\caption{(Co)Annihilation channels into fermions for the Multiplet model.}
\label{fig:Feynmangraphmultiplet3}
\efig

\bfig
\bc
\includegraphics[width=0.95\textwidth]{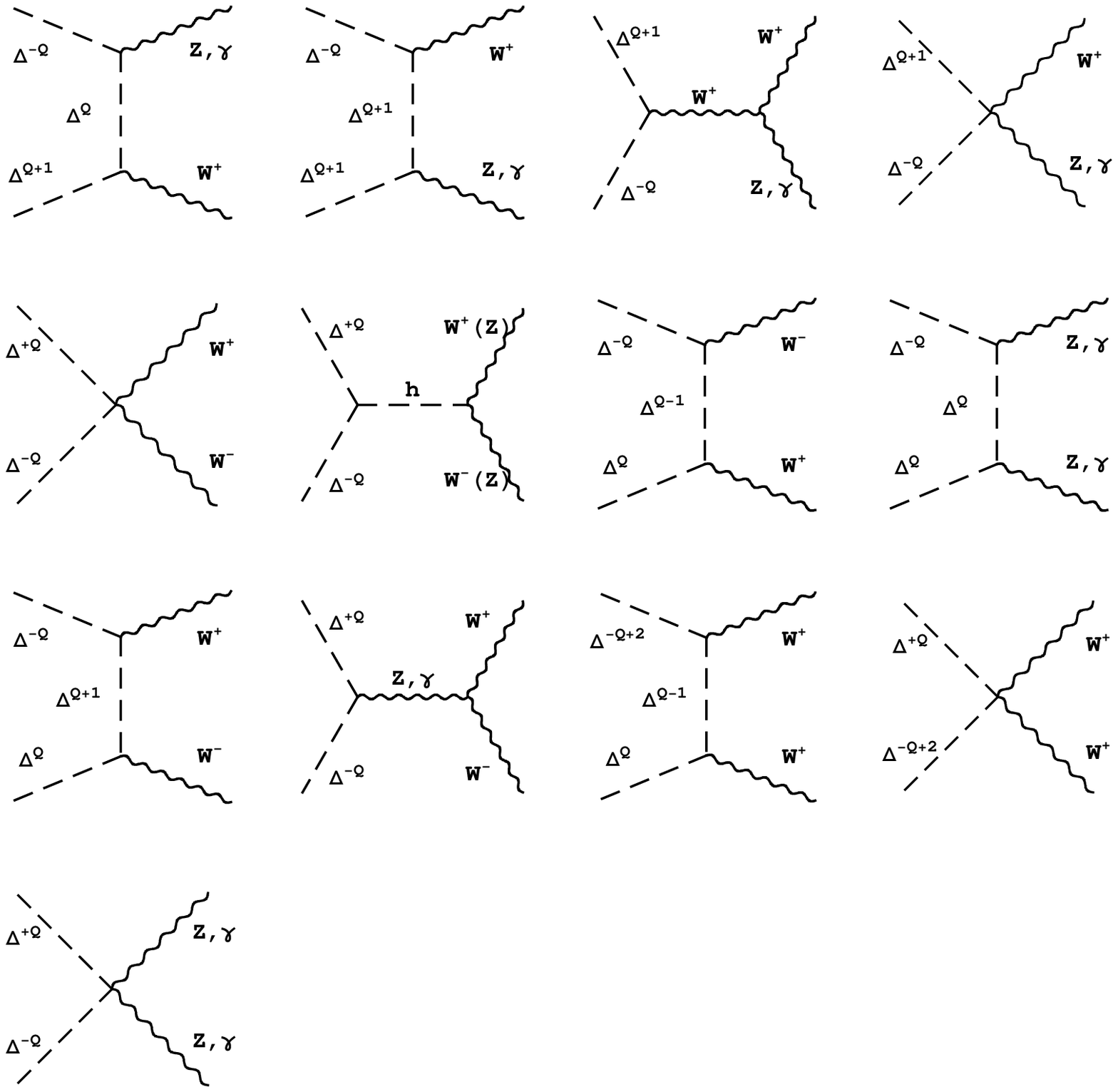}
\caption{Pure gauge (co)annihilation channels for the multiplet model.}
\label{fig:Feynmangraphmultiplet1}
\ec
\efig

\bibliography{scalarDM5}{}
\bibliographystyle{hunsrt}

\end{document}